\documentclass[twocolumn, aps, pra, showpacs, superscriptaddress, floatfix, 10pt]{revtex4-1}

\usepackage{graphicx}
\usepackage{amsmath,amsthm,amsfonts,dsfont}
\usepackage{tabularx}
\usepackage{array}
\usepackage{bbm}
\usepackage{bm}
\usepackage[T1]{fontenc}
\usepackage[colorlinks]{hyperref}
\usepackage{cleveref}

\allowdisplaybreaks

\begin{document}

\title{Experimental Approach to Demonstrating Contextuality for Qudits}

\author{Adel Sohbi}
\email{sohbi@kias.re.kr}
\affiliation{School of Computational Sciences, Korea Institute for Advanced Study, Seoul 02455, Korea}

\author{Ruben Ohana}
\affiliation{Laboratoire de Physique de l'\'Ecole Normale Sup\'erieure, ENS, Universit\'e PSL, CNRS, Sorbonne Universit\'es, 75005 Paris, France}
\affiliation{LightOn, 75002 Paris, France}

\author{Isabelle Zaquine}
\affiliation{LTCI, CNRS, Telecom ParisTech, Univ Paris-Saclay, 75013 Paris, France}

\author{Eleni Diamanti}
\affiliation{LIP6, CNRS, Sorbonne Universit\'es, 75005 Paris, France}

\author{Damian Markham}
\affiliation{LIP6, CNRS, Sorbonne Universit\'es, 75005 Paris, France}
\affiliation{JFLI, CNRS / National Institute of Informatics, University of Tokyo, Tokyo, Japan}

\begin{abstract}
We propose a method to experimentally demonstrate contextuality with a family of tests for qudits.
The experiment we propose uses a qudit encoded in the path of a single photon and its temporal degrees of freedom.
We consider the impact of noise on the effectiveness of these tests, taking the approach of  ontologically faithful non-contextuality.
In this approach, imperfections in the experimental set up must be taken into account in any faithful ontological (classical) model, which limits how much the statistics can deviate
within different contexts.
In this way we bound the precision of the experimental setup under which ontologically faithful non-contextual models can be refuted.
We further consider the noise tolerance through different types of decoherence models on different types of encodings of qudits.
We quantify the effect of the decoherence on the required precision for the experimental setup in order to demonstrate contextuality in this broader sense.
\end{abstract}


\date{\today}
\maketitle

\section{Introduction}

Contextuality is a fundamental property of quantum measurements originally discovered by Bell, Kochen et Specker \cite{Bell:rmp66,KS:jmm67}. It refers to the special property of measurements outcomes statistics to depend on the context in which they are performed, i.e., on the set of measurements that are simultaneously performed. This uses the notion of compatibility, which refers to whether several measurements can be performed without affecting each other that underlies the idea of context and naturally corresponds to a set of measurements that are mutually compatible. Many efforts have been devoted on understanding contextuality \cite{AB:njp11,FLS:qpl13,CSW:arXiv10}. Contextuality has shown an advantage in quantum computing \cite{AB09,raussendorf2013contextuality,HWVE:nat14,DGBR:prx15}, in quantum cryptography \cite{SBKTP09,HHHHPB10}, in quantum state discrimination \cite{SS:prax18} and in self-testing \cite{BRVANCK:prl19}. Moreover, the simulation of quantum contextuality by classical systems requires a larger memory than its equivalent by quantum systems \cite{KGPLC:njp11,FK:njp17,KWB:arXiv18,CGGX:prl18}.

Contextuality generalizes non-locality \cite{FLS:qpl13} and the ability to experimentally observe these two fundamental properties is a key point when it comes to their applicability in information technology tasks. As for non-locality, the most common method to witness contextuality is through inequalities, which uses the statistics of measurements outcomes performed on one or more physical systems. The proof of non-locality with this method is guaranteed by the spatial separation of the physical systems involved, whereas for contextuality all measurements could also be performed onto a single system which makes the proof of contextuality more subtle.

There exist many experimental works regarding the observation of contextuality, generally for qutrit systems as it is the smallest dimension to exihibit contextuality \cite{Bell:rmp66,KS:jmm67}. They use different technologies and different types of encoding using properties of photons such as the polarization and the path \cite{ADTPBC:prl12,AHANBSC:prx13,Ahrens2013,ARBC:prl09}, the  orbital momentum \cite{ACGBXLABSC:pra15}, the path \cite{ZWDCLHYD:prl12,BCAFACCP:pra14,ACGBXLABSC:pra15}, temporal properties \cite{Ahrens2013} or different physical systems such as trapped ions \cite{KZGKGCBR:nat09}, neutrons \cite{BKSSCRH:prl09}, superconducting qubits \cite{JROPMWGWJLF:natcom16}.

Despite all these efforts, when one realizes an experiment it is only possible to perform measurements in their different contexts with a finite precision. This raises the question whether such statistics of measurement outcomes could possibly be performed by classical theory \cite{Meyer:prl99,Kent:prl99,CK:prsla01}. To address this issue, the author in \cite{Winter:pra15}, following the idea in \cite{Spekkens:pra05}, presents an inequality based test to refute any ontological faithful non-contextual model, i.e. a non-contextual model considering the finite precision of the measurement. This model can also be tested through inequalities which can be interpreted as contextuality inequalities with an extra penalty term that takes into account the precision maximum tolerated and the occurrence number of each measurement in different contexts.

There is a deep interest in developping experimental schemes to manipulate qudits and test their quantumness. Indeed using qudits instead of qubits can be beneficial in a range of applications in quantum informnation such as quantum simulation \cite{Neeley722}, quantum algorithms \cite{nphys1150, PhysRevA.94.042307, PhysRevA.96.012306}, quantum error correction \cite{DP:pra13,MSBASJG:prx16,GKWYZ:ieee18}, universal optics-based quantum computation \cite{NCS:prl18}, quantum communication \cite{CDBO:onlinelibrary19, PhysRevLett.123.070505} and fault-tolerant quantum computation \cite{PhysRevX.2.041021,PhysRevLett.113.230501,PhysRevLett.123.070507}, entanglement measurements certification \cite{PhysRevA.100.022117}. From a foundational interest more complex quantum features can be obrtained from higher dimensions such as contextuality \cite{SZDM:pra16} or coherence of measurements \cite{10.1088/1367-2630/abad7e}.

In this work we are interested in unifying an experimental method to observe contextuality for qudits of dimension greater than three and the test of ontological faithful non-contextuality. The authors in \cite{CBCB:prl13} perform a Hardy-like proof of contextuality with a qutrit system, which has been extended to a new family of contextuality inequalities in \cite{SZDM:pra16} for any higher dimension. The qutrit case has also been experimentally observed, in particular in \cite{MANCB:prl14} for which we provide an entension to higher dimension via the new family of contextuality inequalities devlopped in \cite{SZDM:pra16}. We also provide for the experiment realized in \cite{MANCB:prl14} based on \cite{CBCB:prl13} and our proposed setup based on \cite{SZDM:pra16}, the bounds to test any ontological faithful non-contextual model in terms of the precision of the experimental setup used in the proposed experiment.

Moreover, quantum states have to face the effect of decoherence which makes the quantumness of physical system more fragile. For instance in \cite{HCS:njp11}, the authors present a Bell inequality that gives a higher violation for high dimensional systems, but it has been shown that this advantage vanishes when decoherence is taken into account \cite{LBA:pra11,CB:pra11}. Many other works have shown the effects of decoherence on tests of non-locality \cite{LPBZ:pra10,CCA:pra12,SZDM:pra15,BPBBGL:pra15,CAAC:pra}. As the observation of contextuality has been done using different technologies we also consider the noise tolerance of different encodings of the qudit systems for different relevant models of decoherence. We finally gather both of these two approaches to test their cumulative effect on the measurement requirements.

The paper is structured as follows. In section II, we present some preliminary notions; we introduce first an experimental observation of contextuality for a qutrit, then we recall how this contextuality test can be extended to any dimension greater than three and we give the mathematical framework of the ontological faithful non-contextuality inequality. In Section III, we give our method to experimentally observe contextuality with the contextuality inequalities proposed in \cite{SZDM:pra16}. In Section IV, we explain how this experimental method can be used to test ontological faithful non-contextuality. In Section V, we compare the noise tolerance of different encodings of qudits for given decoherence models, namely amplitude and phase damping. Finally in Section VI, we provide a method to quantify how the decoherence affects the required precision to test ontological faithful non-contextuality.

\section{Preliminary Notions}\label{sec:background}
\subsection{Observing Contextuality with the Pentagon Inequality}\label{sec:obscontKCBS}

A graph formalism can be used to represent compatibility and exclusivity relations of dichotomic measurements and to derive non-contextuality inequalities from the graph properties \cite{CSW:arXiv10}.

We consider $N$ dichotomic measurements for which we associate the outcome $X_i$ ($X_i=0$ (`no') or $X_i = 1$ (`yes')) to the vertex $i$ of a graph $\mathcal{G}(V,E)$, where the edges represent the exclusivity and the compatibility of the measurements. Measurements are compatible if it is possible to perform them simultaneously. A context $C$ is a set of pairwise compatible measurements and $C = \{i_1,\dots,i_j \vert (i_k, i_{k'})\in E\}$. Dichotomic measurements are exclusive if they cannot both have an output `yes' simultaneously, $i.e.$ it is not possible that exclusive measurements have the outcome $1$ simultaneously. Hence, $\forall (i,j) \in E$: $P(11\vert i,j) = 0$.
In this framework, a \emph{contextuality inequality} takes the form
\begin{equation}\label{eq:contineq}
\beta = \sum_{i=1} \langle X_i \rangle  \leq \beta_{cl},
\end{equation}
where $\beta_{cl}$ is the classical (non-contextual) bound on the statistics, found by assigning values to $X_i$ in a consistent way across all contexts. Graphically this corresponds to the independence number of the graph (see e.g. \cite{Winter:pra15}). The quantum bound can be higher, and graphically corresponds to the Lovasz function of the graph.

\begin{figure}[tb]
      \centering
      \includegraphics[width=4cm]{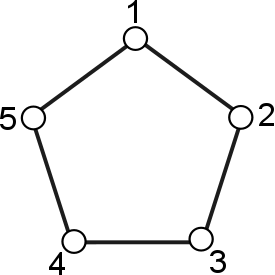}
      \caption{For the KCBS inequality the relations of compatibility and exclusivity are represented by a pentagon.}
    \label{fig:N5}
\end{figure}

The so called Klyachko-Can-Binicio\u{g}lu-Shumovsky (KCBS) inequality \cite{KCBS:prl08} corresponds to a pentagon graph showed in \cref{fig:N5} and can be written in the following form:
\begin{equation}\label{eq:kcbs}
\beta = \sum_{i=1}^5 \langle X_i \rangle  \leq 2,
\end{equation}
where $\{X_i\}$ are the outcomes of the dichotomic measurements that have the compatibility and the exclusivity relations given by the edges of the graph in \cref{fig:N5}.

In the quantum case, rank one projectors can be used to represent the dichotomic measurements. By doing so the maximum quantum violation is known be equal to $\beta_Q = \sum_{i=1}^5 \langle X_i \rangle = \sqrt{5}$ \cite{KCBS:prl08}.

In \cite{MANCB:prl14} the authors present an experimental observation of contextuality on qutrits using the KCBS inequality under the conditions of the logical based proof of contextuality developed in \cite{CBCB:prl13}. In this case, the eigenvectors $\{\vert v_i \rangle\}$ of the five rank one projective measurements $\{P_i = \vert v_i \rangle \langle v_i \vert\}$ they used are:

\begin{align}\label{eq:vectKCBS}
    \vert v_1 \rangle = & \frac{1}{\sqrt{3}}(1,-1,1)^T, \\
    \vert v_2 \rangle = & \frac{1}{\sqrt{2}}(1,1,0)^T, \\
    \vert v_3 \rangle = & (0,0,1)^T,\\
    \vert v_4 \rangle = & (1,0,0)^T, \\
    \vert v_5 \rangle = & \frac{1}{\sqrt{2}}(0,1,1)^T.
\end{align}

The quantum state of the qutrit to be measured is:

\begin{align}\label{eq:contHPstate}
    \vert \eta \rangle = \frac{1}{\sqrt{3}}(1,1,1)^T.
\end{align}

The quantum state is encoded into the path of a single photon.
The computational basis vectors  ($\vert 0 \rangle$, $\vert 1 \rangle$, $\vert 2 \rangle$)  correspond to different distinct paths that the photon can use with $|0 \rangle$ being the upper most path.

To determine a measurement's outcome it is convenient that each outcome can be associated with a specific set of paths.
To do that, we follow the idea presentend in \cite{MANCB:prl14} by rotating the basis to the eigenvector basis of the desired measurement.
In particular in our case, the outcome `1' should be associated with a unique path.
For each vector $\vert v_i \rangle$ we associate a unitary operator $U_i$ which identifies it with the upper path through
\begin{align}\label{eq:unitary}
  & U_i\vert v_i \rangle = \vert 0\rangle.
\end{align}

In the experimental setup, these unitary operators can be implemented by using only beamsplitters.
Note that this equation alone does not fix uniquely $U_i$ and we do not impose that the full basis of paths is assigned to a particular $U_i$. This freedom allows some optimization over the number of beam splitters.

In the pentagon case, each context is composed of two measurements (see \cref{fig:N5}). In order to be able to measure a second projector on the same initial quantum system, the photon is not detected at first but the outcome is stored in a different degree of freedom: the polarization of the photon. The polarization of a photon can be represented in the basis $\{\vert H \rangle, \vert V \rangle \}$, where $\vert H \rangle$ and $\vert V \rangle$ represent horizontal and vertical polarization, respectively. When the photon traverses the path associated with the outcome `1' its polarization is rotated to a vertical position while in the two other paths its polarization remains in a horizontal position. The polarization of the photon becomes entangled with its path. In this way one can keep track of the outcome despite the use of the operation $U_i^\dagger$.

Then the second operation $U_j$ is applied which corresponds to the second measurement of the context, finally the position and the polarization of the photon are measured, which provide the outcome of both projectors of the context. In order to do that we can use polarizing beam splitters followed by the photon detection to give the polarization and the path of the photon. Two detectors are then required for each path.

\begin{figure}[ht]
      \centering
      \includegraphics[width=8.6cm]{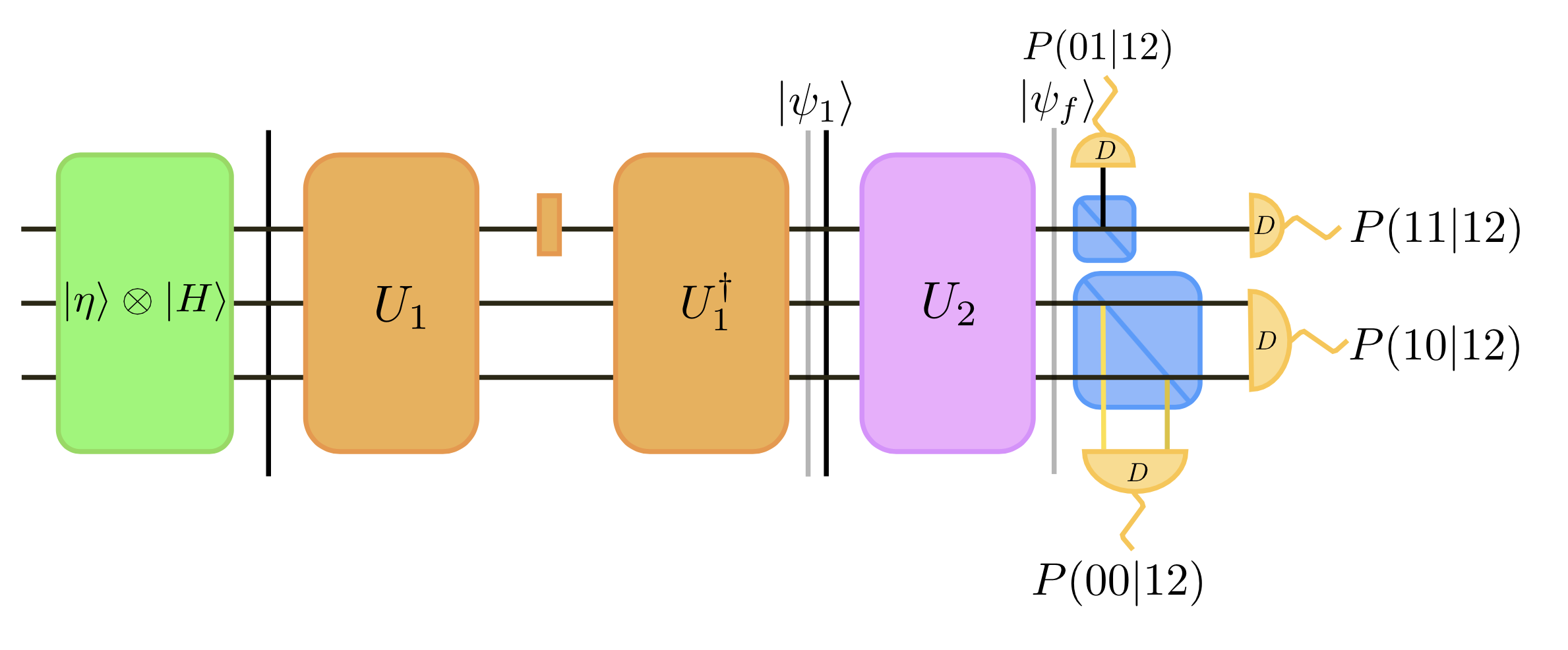}
      \caption{Summary of the measurement process.
The three paths correspond to the basis vectors $\{\vert 0 \rangle, \vert 1 \rangle, \vert 2 \rangle\}$.
After preparing the state $\vert \psi_{ini} \rangle$, we perform $U_1$ and $U_1^\dagger$, associated to outcome  $\vert \psi_1 \rangle$, with a halfwave plate which flips the polarisation in the upper arm in between. Then we proceed to the second measurement by applying the operator $U_2$ giving the state $\vert \psi_f \rangle$. Finally, polarizing beam splitters followed by the photon detection give the polarization and the path. Two detectors are required for each path.}
    \label{fig:contexpN5}
\end{figure}

We describe the evolution of the quantum state step by step, as depicted  in \cref{fig:contexpN5}.
The state $\vert \eta \rangle = \frac{1}{\sqrt{3}}(1,1,1)^T$ is prepared first. Including the polarization degree of freedom, the initial state is
\begin{align}\label{eq:pentexpsiini}
    \vert \psi_{ini} \rangle &= \vert \eta \rangle \otimes \vert H \rangle, \nonumber \\
    & = (\alpha_1 \vert v_1 \rangle  + \alpha_2 \vert v_2 \rangle  + \alpha_{\chi} \vert \chi\rangle) \otimes \vert H \rangle,
\end{align}
where in the second line we expand into the orthogonal states $\{\vert v_1 \rangle, \vert v_2 \rangle , \vert \chi \rangle \}$ which represents the context with the vertices $\{1,2\}$ from the the pentagon.

Next, $U_1$ is applied, which encodes the vector $|v_1\rangle$ onto the upper path (see \cref{eq:unitary}).  The encoding of the polarisation flip onto the upper path is done via a half wave plate. The corresponding unitary is a control flip (a flip in polarisation contolled by the path). This is followed  by $U_1^\dagger$.
In this way, the quantum state between $U_1^\dagger$ and $U_2$ in \cref{fig:contexpN5} is
\begin{align}\label{eq:pentexpui2}
\vert \psi_{1} \rangle = \alpha_1 \vert v_1 \rangle \otimes \vert V \rangle + (\alpha_2 \vert v_2 \rangle + \alpha_{\chi} \vert \chi\rangle) \otimes \vert H \rangle.
\end{align}

After the operation $U_2$ in \cref{fig:contexpN5} the quantum state is
\begin{align}\label{eq:pentexpf}
    \vert \psi_f \rangle = &  \alpha_2 |0 \rangle  \otimes \vert H \rangle + \alpha_1 U_2 \vert v_1 \rangle \otimes \vert V \rangle + \alpha_\chi U_2 \vert \chi\rangle \otimes \vert H \rangle,
\end{align}
where the first term corresponds to the outcome $X_2=1$ and $X_1=0$, the second term corresponds to the outcome $X_2=0$ and $X_1=1$ and the last term corresponds to the outcome $X_1=X_2=0$.

The number of beam splitters needed varies for each unitary in $\{ U_i \}_{i \in V}$. The matrices of unitary operators are then described by the product of the matrices of each used beam spitters. They are presented in \cref{sec:appuniN5}.

By following this procedure for the full set of unitary operators corresponding to all contexts, \cite{MANCB:prl14} obtains
\begin{equation}
    \beta_{Q,exp} = 2.078 \pm 0.038,
\end{equation}
which demonstrates an experimental violation of the KCBS inequality.

\subsection{Graph Family for an Extension of the KCBS Inequality}\label{sec:KCBSExtention}

In \cite{SZDM:pra16}, a family of graphs is developed, associated to a family of contextuality inequalities that gives an extension to the KCBS inequality and logical based proof of contextuality in \cite{CBCB:prl13} . The graph construction goes as follows.
For $|V|=N\geq 5$, we define a $G=(V,E)$, such that it contains two complete (i.e edges between all nodes) subgraphs $G_A$ and $G_B$ such that:
\begin{itemize}
    \item $V_A=\{2,\dots,(N+1)/2\}$ and $V_B=\{(N+1)/2+1,\dots,N\}$ for $N$ odd.
    \item $V_A=\{2,\dots,N/2+1\}$ and $V_B=\{N+/2+1,\dots,N\}$ for $N$ even.
    \item The vertices $2$ and $N$ are connected.
    \item The vertex $1$ is connected to all vertices except $2$ and $N$.
\end{itemize}

\begin{figure}[ht]
      \centering
      \includegraphics[width=5cm]{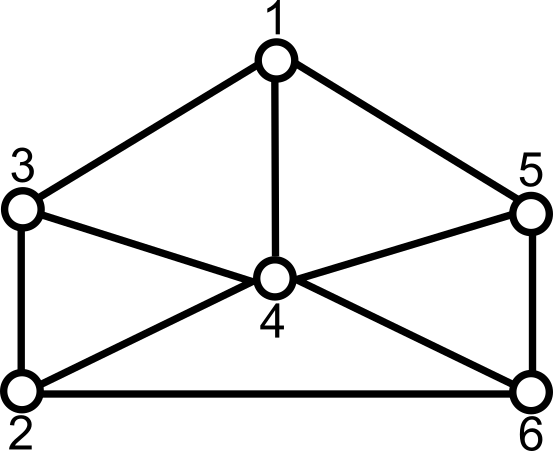}
      \caption{The graph construction for odd $N$ ($N=6$). The two complete subgraphs $G_A$ and $G_B$ are formed respectively by the set of vertices $V_A=\{2,3,4\}$ and $V_B=\{4,5,6\}$.}
    \label{fig:N6}
\end{figure}

\begin{figure}[ht]
      \centering
      \includegraphics[width=5cm]{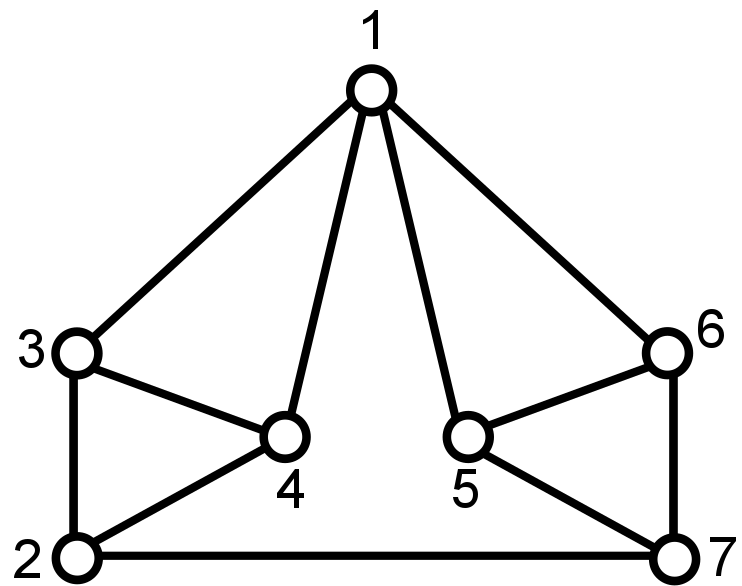}
      \caption{The graph construction for odd $N$ ($N=7$). The two complete subgraphs $G_A$ and $G_B$ are formed respectively by the set of vertices $V_A=\{2,3,4\}$ and $V_B=\{5,6,7\}$.}
    \label{fig:N7}
\end{figure}

The two subgraphs can share a vertex, if $N$ is even $V_A \cap V_B = \{N+/2+1\}$.
The graph represented in \cref{fig:N6} is an even case: $N=6$, in \cref{fig:N7} an odd case: $N=7$.

In particular for the case $N=6$, there are five different contexts:
\begin{itemize}
    \item $C_1 = \{i \in \{2,3,4\}\}$,
    \item $C_2 = \{i \in \{1,3,4\}\}$,
    \item $C_3 = \{i \in \{4,5,6\}\}$,
    \item $C_4 = \{i \in \{1,4,5\}\}$,
    \item $C_5 = \{i \in \{2,4,6\}\}$,
\end{itemize}
where each context corresponds to a complete subgraph in \cref{fig:N6} composed by three vertices leading to the three measurements per context.

The contextuality inequality  is:

\begin{align}\label{eq:kcbsN}
\sum_{i=1}^N \langle X_i \rangle\leq 2.
\end{align}

In \cite{SZDM:pra16} it is shown that there exist a quantum state and a set of measurements for each graph such that $\beta_Q = 2 + \frac{1}{9}$ which provides a violation of inequality \cref{eq:kcbsN}.

\subsection{Ontologically Faithful Non-Contextuality}\label{sec:ONFC}

In order to make an experimental observation of contextuality it is needed to perform the same measurement in its different contexts. Unfortunately, it is only possible to ensure with a finite precision $\epsilon$ that a specific measurement is performed the same in all its contexts when taking into consideration the experimental limitations. If a measurement in its different contexts is treated differently by a classical theory by assigning different random variables depending on the context, it is then possible to simulate its outcome statistics with such a classical theory \cite{Meyer:prl99,Kent:prl99,CK:prsla01,BK:shpspb01,Winter:pra15}. The idea behind \cite{Winter:pra15} is that a classical theory that would treat the same measurement differently in its different contexts also needs to follow some constraints that capture the precision $\epsilon$; such a classical theory is called $\epsilon$-ontologically faithful non-contextual.

In the classical case, the \emph{$\epsilon$-ontologically faithful non-contextual} model is as follows:
\begin{itemize}
    \item For any context $C$ : $\sum_{i\in C} \langle X_{i,C} \rangle  \leq 1$.
    \item For all measurements in contexts $C$ and $C'$ associated with the vertex $i\in V$ : $\text{Prob}(X_{i,C} \neq X_{i,C'}) \leq \epsilon$.
\end{itemize}
The probability that the result of a measurement is different in two different contexts must be below $\epsilon$.

In the quantum case, the result of a projective measurement $P_i$ is obtained by measuring a projector $P_{i,C}$ giving the result $X_{i,C}$ for each context $C$. Reference \cite{Winter:pra15} introduces a quantum model with a finite precision $\epsilon$ as the follows:
\begin{itemize}
    \item For any context $C$ : $\sum_{i\in C}  \langle X_{i,C} \rangle \leq 1$.
    \item For all measurements in a context $C$ associated with the vertex $i\in V$ : $\vert \vert P_i - P_{i,C} \vert \vert \leq \epsilon$, where $\vert \vert P_i - P_{i,C} \vert \vert$ is the norm of the distance defined by:
    \begin{equation}\label{eq:distONC}
        \vert \vert A - B \vert \vert = \underset{\rho}{\max} \vert Tr (\rho (A- B))\vert,
    \end{equation}
    where $A$ and $B$ are operators and $\rho$ is a density matrix.
\end{itemize}

In other words, this model ensures that the distance between a theoretical projection and any of its experimental realizations is not greater than $\epsilon$.

From \cite{Winter:pra15}, if one has a contextuality inequality of the form $\beta = \sum_i \langle X_i \rangle \leq \beta_{cl}$ as before, for an $\epsilon$-ontologically faithful non-contextual model, the inequality becomes
\begin{equation}\label{eq:szdmineq}
    \beta = \sum_i \langle X_i \rangle \leq \beta_{cl} + \epsilon \sum_i (k_i -1),
\end{equation}
where $k_i$ is the number of contexts in which the results associated with the vertex $i$ appears.

For a quantum violation of $\beta_Q$,  the inequality can be rewritten
\begin{equation}\label{eq:ineepsiONC}
    \epsilon \geq \frac{\beta_Q - \beta_{cl}}{\sum_i (k_i - 1)}.
\end{equation}
A physical experiment that can violate this inequality can not be described by an $\epsilon$-ontological faithful non-contextual model.

\section{Proposed Qudit Experiment}\label{proposedexp}

Compared to the pentagon, the family of graphs developed in \cite{SZDM:pra16} may have contexts with more than two measurements, hence polarization is not enough to store the outcomes of each intermediate measurement if one wants to use the experimental setup in \cite{MANCB:prl14}. Instead of using the polarization of the single photon we propose to use its temporal properties. Hence, instead of using a wave plate to change the polarization of the photon we propose to use a delay $\Delta t_i$ corresponding to a measurement that could be applied to the photon. Each measurement of a context (except the last one for which the outcome is known only from the path) has its own characteristic time delay to be able to recognize which path the photon has used. Experimentally, the time delay of a photon can be known if the experimental setup uses a deterministic single photon source or a photon pair where one of the photons is used for the contextuality test and the other one as a time reference.

The initial quantum state, $\vert \psi_{ini} \rangle$, is
\begin{align}\label{eq:graphexpsiini}
    \vert \psi_{ini} \rangle &=\vert \eta \rangle \otimes \vert 0 \rangle_t,\\
&= (\sum_{i\in C} \alpha_i |v_i\rangle + \alpha_{\chi_C}|\chi_C\rangle)  \otimes \vert 0 \rangle_t,
\end{align}
where $\vert \eta \rangle$ is the quantum state used to demonstrate contextuality,  $\vert 0 \rangle_t$ corresponds to zero time delay and where we have expanded
for a given context C in the associated vectors.

In this case we encode onto time delay instead of the polarization, which similarly corresponds to a control unitary - a time delay controlled by the path.
Employing the same strategy of sandwiching this control delay between the unitary operators encoding the measurement basis, with delay $\Delta_{t_i}$ for $U_i$, we get the final state before measurement
\begin{align}\label{eq:contexpfinstategraphf}
   \vert \psi_f \rangle & = \alpha_{i_f} \vert v_{i_f} \rangle \otimes \vert 0 \rangle_t + \sum_{i\in C, i\neq i_f} \alpha_i U_{i_f}\vert v_i \rangle \otimes \vert \Delta t_i \rangle_t \notag\\
   &+  \alpha_{\chi_C} U_{i_f} |\chi_C\rangle  \otimes \vert 0 \rangle_t,
\end{align}
where $U_{i_f}$ is the unitary operator of the last measurement and $\vert v_{i_f} \rangle$ is the eigenvector with eigenvalue 1 of the projector of the last measurement. The first term in \cref{eq:contexpfinstategraphf} corresponds to the case where the outcome will be $X_{i_f}=1$ and $X_{i\in C/i_f}=0$, the second term corresponds to $X_{i_f}=0$ and there exist $j\in C/ i_f$ such that $X_j=1$ and $X_{i\in C/\{i_f,j\}}=0$ and the last term corresponds to $X_{i\in C}=0$.

\begin{figure}[ht]
      \centering
      \includegraphics[width=8.6cm]{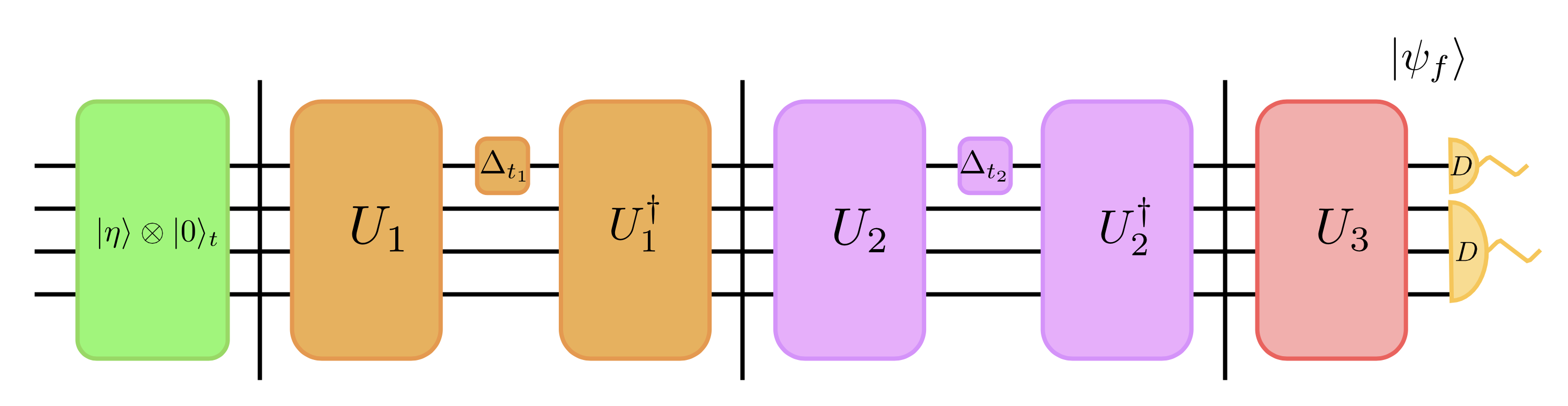}
      \caption{Proposed setup for a context of three measurement which corresponds to the case $N=6$. The initial state is prepared in the state $\vert \psi_{ini} \rangle$. The detection of the path and the delay of the single photon give the outcomes of the measurements.}
    \label{fig:contexPN6}
\end{figure}

In \cref{fig:contexPN6} there is a summary of the process in the case $N=6$ where each context is composed by three measurements.

This process allows also to check the exclusivity conditions between the measurements because if the delay obtained for the single photon does not correspond to a single $\Delta t_i$ but a combination of different $\Delta t_i$ then the exclusivity relations imposed by the graph are not satisfied. The compatibility conditions can be validated by checking that the statistics of the measurements is invariant by the order in which the blocs of unitary operations are applied. Moreover the values of the $\Delta t_i$ need to be taken such that none of the time delays is a sum of other time delays. Hence $\forall i \in V$:

\begin{equation}\label{eq:time-delay}
\Delta t_i \neq \sum_{k \neq i} \lambda_{k,i} \Delta t_k,
\end{equation}
where $\lambda_{k,i} \in \{0,1\}$. This ensures we can keep track of the outcome of each measurement in the context and experimentally verify the exclusivity.

\section{Refuting Ontologically Faithful Non-Contextuality}\label{sec:testOFNC}

In this section we derive minimum bounds on the precision required in order to escape the possibility of an ontologically faithful non-contextual model, that is to demonstrate contextuality in this broader sense.
We present what this means in terms of matrices of the beam splitters. We do this explicitly for the KCBS and N=6 cases, similar studies can be easily made for larger graphs.

\subsection{Testing with the KCBS Inequality}

Earlier, in \cref{sec:obscontKCBS}, we presented an example of an experimental observation of contextuality that uses the KCBS inequality \cite{MANCB:prl14} under the conditions of the paradox developed in \cite{CBCB:prl13}.

An inequality can be used to test the ontologically faithful non-contextuality to overcome the impossibility to measure perfectly the same observable in its different contexts (see \cite{Winter:pra15} and \cref{sec:ONFC}).

To see how this works in our case, we assume that the finite precision $\epsilon$ used in ontologically faithful non-contextuality in \cite{Winter:pra15} is due to an imperfection of the beam splitters used in the experiment to build the unitary operators in \cite{MANCB:prl14}. For simplicity, we consider the same typical error denoted $\delta$ for all beam splitters (in absolute value). In \cref{app:KCSBnoisyUi}, we present the unitary operators where the factor $\delta$ has been introduced in each beam splitter. We neglect the phase errors on the reflectance and transmission.

In the case of the pentagon, the necessary condition to verify ontologically faithful non-contextuality is $\epsilon <\frac{1}{45}$ \cite{Winter:pra15}, where $\epsilon$ is an upper bound on the distance between the projection $P_i=\vert v_i \rangle \langle v_i \vert$ and its experimental realization in different contexts. Experimentally, each projector $P_i$ is realized thanks to the unitary operation $U_i$. We define $P_{i,\delta} = \vert v_{i,\delta} \rangle \langle v_{i,\delta} \vert$ and $U_{i,\delta} \vert v_{i,\delta} \rangle = \vert 0 \rangle$ the projector and the unitary operation when the typical error $\delta$ is considered. Hence,

\begin{align}\label{eq:delta}
    \Delta_i & = \vert \vert P_i - P_{i,\delta} \vert \vert < \epsilon, \notag\\
    \Delta_i & = \vert \vert \vert v_{i} \rangle \langle v_{i} \vert - U_{i,\delta}^\dagger\vert 0 \rangle \langle 0 \vert U_{i,\delta} \vert \vert < \epsilon,
\end{align}
where the states in the first term are given in \cref{eq:vectKCBS} and the second term can be calculated by with the unitary operators given in \cref{app:KCSBnoisyUi}.
The norm of the matrix $M_{i,\delta}$ is:

\begin{align}\label{eq:contdeltaepsN5}
    \vert \vert M_{i,\delta} \vert \vert = & \max \vert \sigma_i(\delta) \vert,
\end{align}
where $\sigma_i(\delta)$ are the singular values of $M_{i,\delta}$.

In the case of the KCBS inequality, if we have maximum quantum violation, for it to correspond to an ontologically faithful non-contextual model we would require $\max\vert \sigma_i(\delta)\vert<\frac{1}{45}, \forall i\in V$.

\begin{figure}[ht]
    \centering
    \includegraphics[width=9cm]{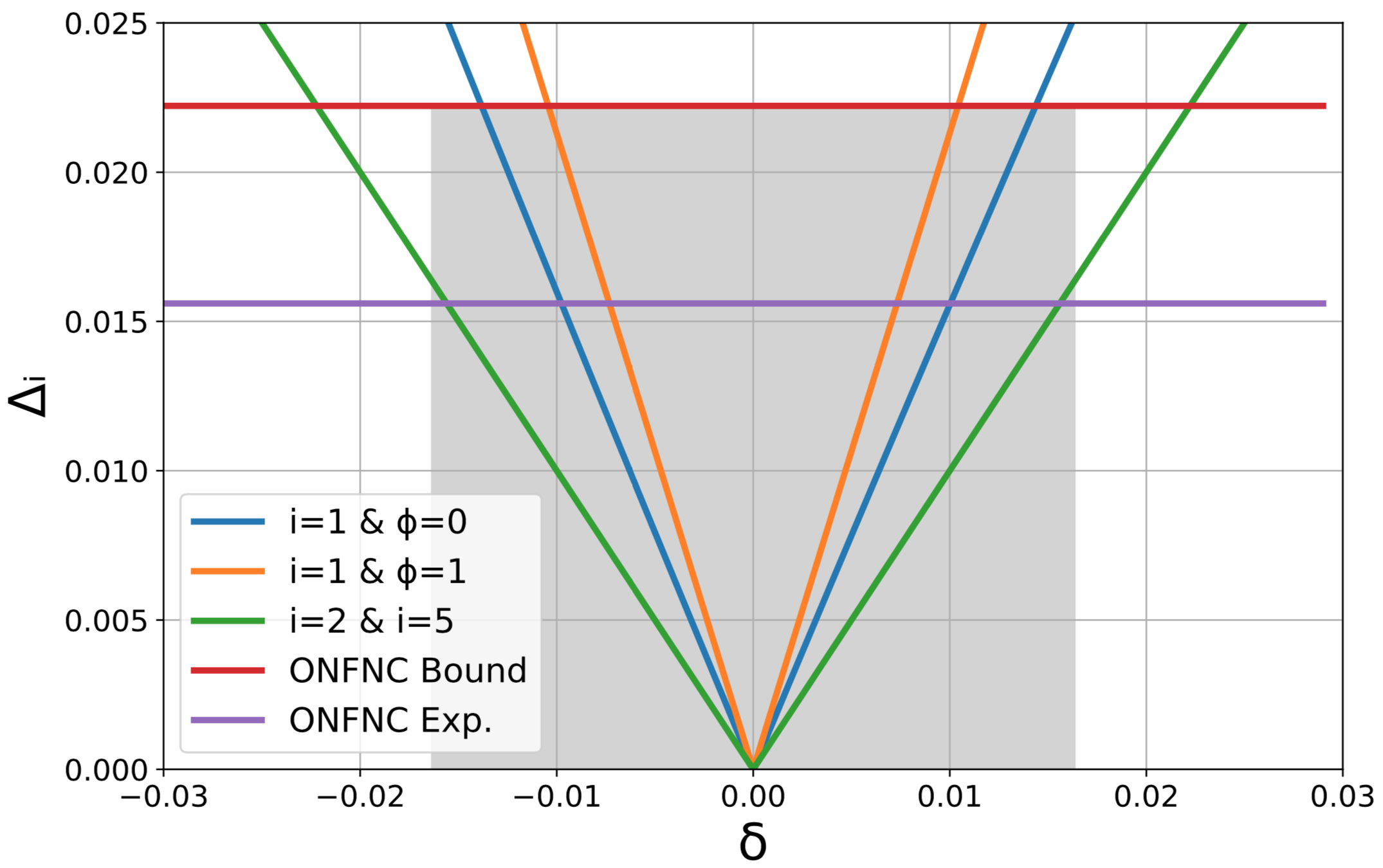}
    \caption{Distance $\Delta_i$ between theoretical and the expected experimental operators as a function of $\delta$. The ontologically faithful non-contextuality bound is $ \frac{1}{45}$ and is represented by the upper horizontal line. The other curves correspond to the largest eigenvalues of the unitary $U_1$ (represented by two curves $\sigma_{1,\phi = \{0,1\}}$) and $U_2$ (which is also $U_5$ because the beam splitters are the same) which are functions of $\delta$. The grey area represents the values of $\delta$ leading to a violation. The lower horizontal line corresponds to the ontologically faithful non-contextuality bound with the experimentally obtained violation in \cite{MANCB:prl14}.}
  \label{fig:ONFNC_N5}
\end{figure}

In \cref{fig:ONFNC_N5} we show the curves of $\Delta_i$ of the different unitary operations, in order to compare with the value of $\epsilon$ represented by the upper horizontal line. \cref{eq:ineepsiONC} is violated, hence refuting any $\epsilon$ ontologically faithful non contextuality only in the range of parameter $\delta$ where the curves representing $\Delta_i$ are below the upper line representing the $\epsilon$ bound. Because the implementation of the unitary operators $U_3$ and $U_4$ (\cref{app:KCSBnoisyUi}) do not require the use of any beam splitters (but requires to relabel the paths instead) they are not shown in the figure. Moreover, because $U_2$ and $U_5$  are very similar (\cref{app:KCSBnoisyUi}) we have $\Delta_2 =\Delta_5 $, so they are shown as one in the \cref{fig:ONFNC_N5}. Finally, $U_1$ requires the use of an additional factor called $\phi$ (\cref{app:KCSBnoisyUi}); this is because this unitary operator is composed of two beam splitter operators and while both have the same error $\delta$ we consider the case where one has error $\delta$ and the other one has $\pm\delta$. Additionally, we added a lower horizontal line which corresponds to the ontologically faithful non-contextuality bound with the experimentally obtained violation in \cite{MANCB:prl14}. This value correspond of the value of $\epsilon$ when the \cref{eq:ineepsiONC} is saturated for the experimenlly obtained violation in \cite{MANCB:prl14}.

The maximum value of $\delta$ for a violation is then the maximal $\delta$ for which all curves are below the horizontal line. Thus we get the value $\delta_{th}=\pm 0.0164974$. This corresponds to a maximum tolerated error of $1.6\%$ of the coefficient of transmission and reflection of the beam splitters.

\subsection{Testing with the Extension of the KCBS Inequality for $N=6$}

This method can be extended to other inequalities. We propose to study the particular case of the graph $N=6$ (see \cref{fig:N6}) of the inequalities developed in \cite{SZDM:pra16} with the experiment we propose. In this case ($d=4$), the quantum state and the eigenvectors of rank one projective measurements are given in \cite{SZDM:pra16}:

\begin{align}\label{eq:Mimpair}
\vert \eta \rangle & = \frac{1}{\sqrt{6}}(\sqrt{2}, 1, 1, \sqrt{2})^T,\\
\vert v_1 \rangle & = \frac{1}{\sqrt{6}}(-\sqrt{2}, 1, 1, -\sqrt{2})^T,\\
\vert v_2 \rangle & = (1, 0, 0, 0)^T,\\
\vert v_3 \rangle & = \frac{1}{2}(0,1,1,\sqrt{2})^T,\\
\vert v_4 \rangle & = \frac{1}{\sqrt{2}}(0, -1, 1, 0)^T,\\
\vert v_5 \rangle & = \frac{1}{2}(\sqrt{2},1,1,0)^T,\\
\vert v_6 \rangle & = (0, 0, 0, 1)^T.
\end{align}

By computing $\beta_Q = \sum_{i=1}^6 \vert \langle v_i \vert \psi \rangle \vert^2$ one can derive the value $\beta_Q = 2 + \frac{1}{9}$ which violates the inequality shown in \cref{eq:szdmineq}.

To obtain the unitary operators $\{U_i\}_{i \in V}$ as $U_i \vert v_i \rangle = \vert 0 \rangle$ (see \cref{eq:unitary}), the following technique is applied. The first step is to count the number of nonzero components of each $\vert v_i \rangle$ in the path basis, to determine the number of beam splitters. By adjusting the coefficients of reflection and transmission of a beam splitter, it is possible to cancel an unwanted component of the vector $U_i\vert v_i \rangle$ until a single non-zero component to the desired path is obtained. In this way we can obtain the unitary  \cref{eq:unitary} as a circuit of beam splitters. The number of beam splitters required is the number of unwanted components and is optimum. $U_1$ requires three beam splitters, two for $U_3$ and $U_5$, one for $U_4$ and zero for $U_2$ and $U_6$. The unitary operators for $N=6$ are giving in the \cref{app:N6Th}.

To test ontologically faithful non-contextuality with the extension of KCBS inequality it is required that the distance between the projection $P_i$ and the corresponding projections measured in each context $P_{i,\delta}$ is less than $\frac{1}{9(N + 3)} = \frac{1}{81}$ \cite{SZDM:pra16}. Following the same process described in the previous section we can compute all the $\Delta_i$ from the \cref{eq:delta,eq:Mimpair} and the unitary operations described in \cref{app:N6Th}.

\begin{figure}[ht]
    \centering
    \includegraphics[width=7.6cm]{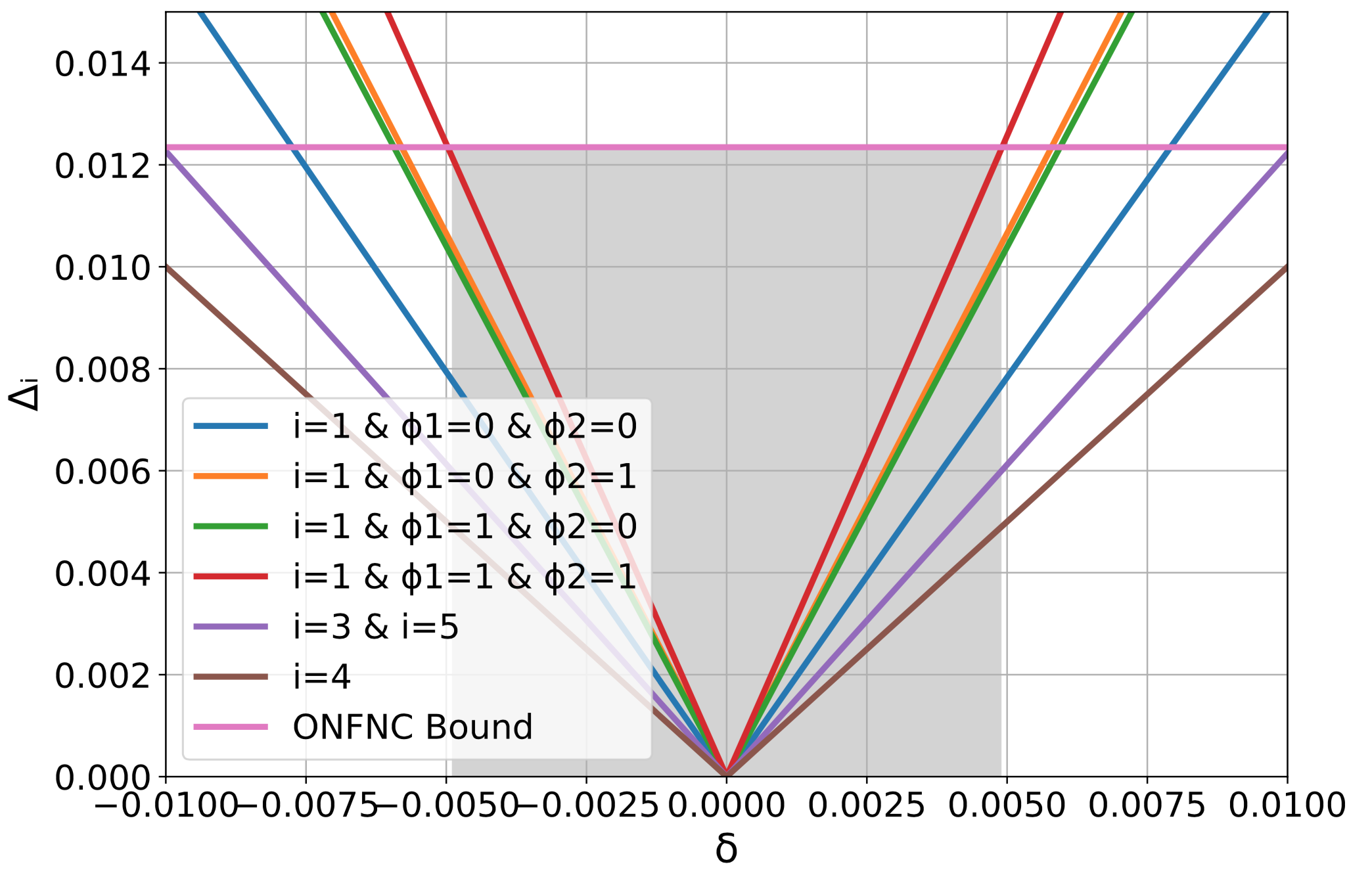}
    \caption{Distance between ideal and the expected experimental operators depending on $\delta$. The ontologically faithful non-contextuality bound is $ \frac{1}{81}$ and is represented by a horizontal line. The other curves correspond to larger values of the different unit operations $ U_1 $ (for every combination of $\phi_1$ and $\phi_2$) $U_3$ (with $U_5$) and $U_4$ are functions of $\delta$, $U_2$ and $U_6$ are not shown as they don't need beam splitters. The grey area represents the values of $\delta$ leading to a violation.}
  \label{fig:ONFNC_N6}
\end{figure}

In \cref{fig:ONFNC_N6}, the different curves represent the maximum values of $\sigma_i(\delta)$ of each unitary operators built with beam splitters. As the optical networks to build $U_2$ and $U_6$ have no beam splitters they do not appear in the figure. The horizontal line is the threshold above which there is no observable violation. The upper-bound to violate is the value of $\delta$ for which all curves are below the horizontal line $\epsilon = \frac{1}{81}$. Thus we obtain the value $\delta_{th} = 0.0049$. This corresponds to a maximum tolerated error of $ 0.49 \% $ of the coefficients of reflection and transmission for each beam splitter. The tolerance is smaller than what was obtained for the pentagon.

While the violation of the contextual inequality seems constant for all graphs \cite{SZDM:pra16}, it seems that the ontologically faithful non-contextuality is becoming more difficult to refute when the number of vertices increases. This can be understood by the fact that the largest dimensions require more elements in the experimental setup as resulting in a greater accumulation of errors.

\section{Decoherence Effects on the Extension of the KCBS Inequality}\label{sec:decKCBSmain}
\subsection{Robustness of the KCBS Inequality against Decoherence}\label{sec:decKCBS}

In our experimental proposal, the main noisy effects would be the errors due to the imperfection of the beam splitters, and photon loss.
We have treated the first case in the previous section and in these experiments we would effectively be post selecting onto the no-photon-loss case.
However, we are also interested in investigating the case of other noise models for further potential implementations.
We therefore consider in this section the experimental limitation for various decoherence models.
In the spirit of \cite{SZDM:pra15}, we consider two decoherence models: amplitude and phase damping and we compute the threshold, $i.e$ the maximum tolerated noise until the violation is lost in each case. For that we use the Kraus operators formalism for these two models for any qudit of dimension $d$:

\begin{itemize}
    \item For amplitude damping:
    \begin{equation}\label{eq:Krausampqudit}
    A_k = \sum_{r=k}^{d-1}\sqrt{\binom {r}{k}}\sqrt{(1-\gamma)^{r-\gamma}\gamma^k}\vert r - k \rangle \langle r \vert,
    \end{equation}
    where $k$ is a positive integer verifying $ k \leq d - 1$ and $\gamma \in [0,1]$ is the noise factor.

    \item For phase damping:
    \begin{equation}\label{eq:Krausphqudit}
        P_k =
        \left\{
        \begin{split}
            \sum_{r = 0}^{d-1}(1 - \lambda)^{r^2/2} \vert r \rangle \langle r \vert \text{ si }k=0,\\
            \sqrt{(1 - (1 - \lambda)^{r^2})} \vert r \rangle \langle r \vert \text{ sinon,}
        \end{split}
        \right.
    \end{equation}
    where $k$ is a positive integer verifying $k \leq d - 1$ and $\lambda \in [0,1]$ is the noise factor.
\end{itemize}

In the first instance we are interested to know the threshold on the noise factors, $i.e.$ the value of $\lambda$ or $\gamma$ above which the inequalities in Eq.~\ref{eq:kcbsN} are not violated anymore.

Moreover, we apply these decoherence models into three different types of encoding of qudits:

\begin{itemize}
    \item \emph{Single qudit}. We apply the Kraus operators as they are presented in \cref{eq:Krausampqudit} and in \cref{eq:Krausphqudit}. In this case the noise is applied onto the single qudit quantum system.

    \item \emph{System composed by qubits}. The use of qubits imposes restriction on the possible dimension of the qudit. In fact, the dimension of the qudit has to match with the product of the different subsystems, i.e. the dimension $d$ of the qudit has to be equal to $2^n$, where $n$ is the number of qubits in the composite system. In this case the Kraus operators are applied on each subsystems.

    \item \emph{State composed by a symmetric state}. In this case the qudit is a composite system of qubits forming a symmetric state, where the dimension of the qudit matches with the dimension of the symmetric subspace of the qubits, i.e. the dimension of the qudit is equal to $n+1$ for $n$ qubits forming a symmetric state. The noise is then applied to each qubits separately via the Kraus operators.
\end{itemize}

\begin{figure}[ht]
      \centering
      \includegraphics[width=9.2cm]{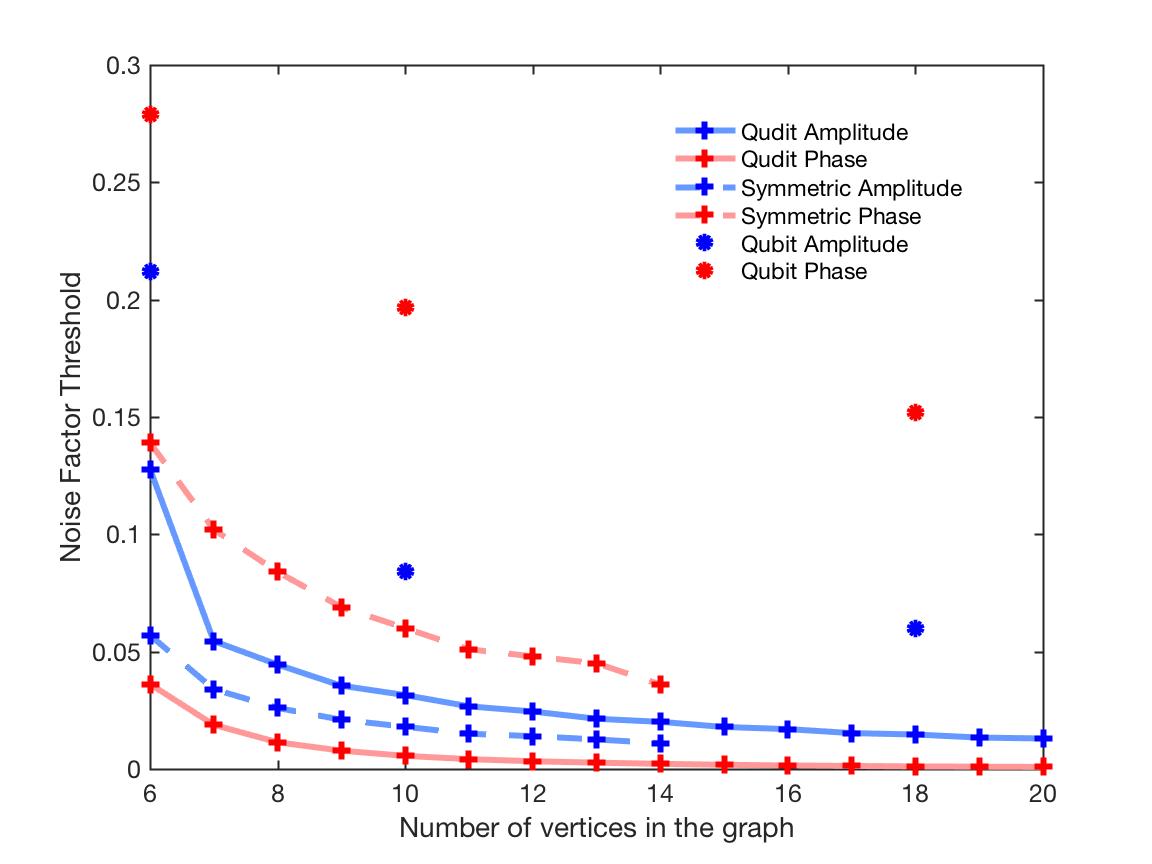}
      \caption{Noise factor threshold for amplitude and phase damping ($\gamma$ and $\lambda$) for the three types of encoding.}
    \label{fig:decont}
\end{figure}

In \cref{fig:decont}, we show that the threshold on the noise factor decreases when the number of vertices in the graph increases for all types of noise and types of encoding. The use of a composite system of qubits seems more advantageous for both types of noises. Phase damping seems to have a smaller effect onto the qubits composite systems and the symmetric states, whereas amplitude damping has a smaller effect onto the unique qudit system. This behavior might not be a general trend and could depend on the inequality that is considered.

It is actually non-trivial to compare the different encoding systems. However, is it possible to understand why there are such differences. This is because a noise model will act differently on different encodings. For instance, for a $d=4$ qudit, we can decide to encode it into the indistinguishable photon number basis $\{\vert 0 \rangle, \vert 1 \rangle, \vert 2 \rangle, \vert 3 \rangle\}$ or with two distinguishable photon basis $\{\vert 00 \rangle,\vert 01 \rangle,\vert 10 \rangle, \vert 11 \rangle\}$. If one applies an amplitude damping model on the third vector of each basis $\vert 2 \rangle$ gives $\vert 1 \rangle$, that is the second vector of the basis, whereas in the second case $\vert 10 \rangle$ gives $\vert 00 \rangle$, which results in the first vector of the basis. This effect will affect the outcome of the measurement, hence the possible violation.

\subsection{Effects of the Decoherence on Ontological Faithful Non-Contextuality Inequalities}\label{sec:decONFCKCBS}

\begin{figure}[ht]
      \centering
      \includegraphics[width=8.5cm]{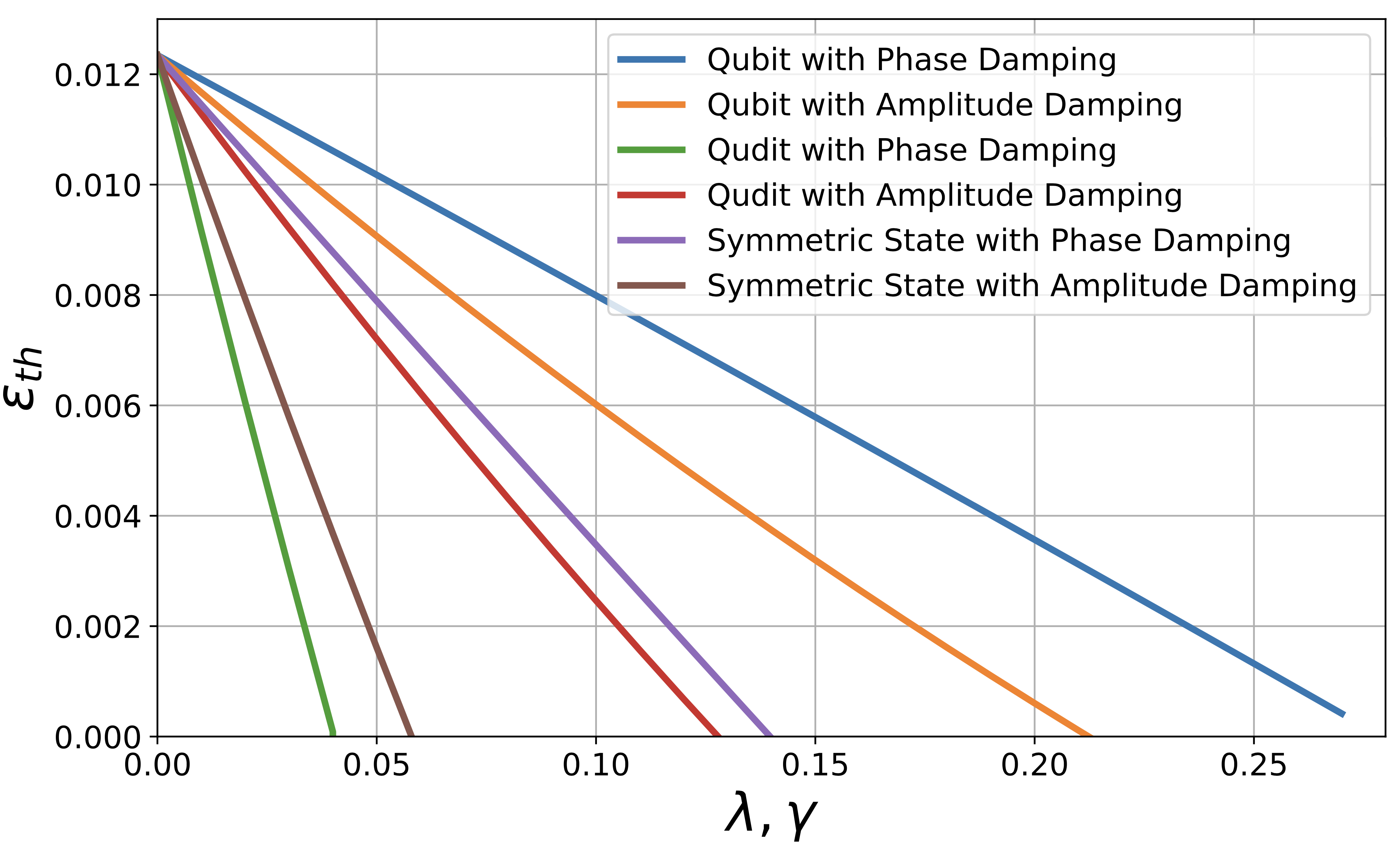}
      \caption{Threshold on the error $\epsilon_{th}$ as a function of the noise parameter $\lambda$ of $\gamma$ for $N=6$.}
    \label{fig:ONFNC_N6_Dec}
\end{figure}

In \cref{sec:decKCBS}, we have seen only the threshold of tolerated noise, $i.e.$ the maximum value of noise above which we do not observe the violation of the inequality. However, this arises from a continuous process where the noise gradually increases and the violation decreases consequently. During this transition, as the violation decreases, the value of $\epsilon$ reduces accordingly because $\epsilon$ is proportional to the difference between the quantum violation and the classical bound. We define $\epsilon_{th}$, the value of $\epsilon$ that saturates \cref{eq:ineepsiONC}. It depends on the noise parameters $\lambda$ or $\gamma$ because $\beta_Q$ is dependent on the noise. As $\beta_Q$ decreases when $\lambda$ or $\gamma$ increases, $\epsilon_{th}$ decreases as well. This is shown in \cref{fig:ONFNC_N6_Dec} where the values $\epsilon_{th}$ are plotted as a function of the noise parameters $\lambda$ and $\gamma$ for the different encoding and decoherence models used in \cref{fig:decont}. In these process we consider the beam splitters without imperfection. Hence, these values correspond to where the threshold shown as the horizontal bar in \cref{fig:ONFNC_N5} (the upper one) and \cref{fig:ONFNC_N6} would be when taking into account both damping models with the beam splitters imperfections at the same time.

With this information, one can assess the practicality of being able to demonstrate non-contextuality against any ontologically faithful model.

In Fig.\ref{fig:ONFNC_N5}, the lower horizontal line corresponds to the ontologically faithful non-contextuality bound with the experimentally obtained violation in \cite{MANCB:prl14}. In this case, as the violation is lower, the threshold on the precision is $\delta_{th}=\pm 0.0116$. This corresponds to a maximum tolerated error of $1.16\%$ on the coefficients of the reflection and transmission for all beam splitters.

\section{Conclusion}

In this article we address the challenges of an experimental observation of contextuality, in particular through the ontologically faithful non-contextuality model described in \cite{Winter:pra15}. We propose an experiment which can theoretically be used for any arbitrary dimension of qudit with an arbitrary number of successive measurements. This is possible by encoding the qudit in a path of a single photon and its temporal properties by using a control-time-delay operation. Considering specific experimental limitations enables the derivation of an ontologically faithful non-contextuality in terms of practical factors such as the reflectivity of the beam splitters in this case. We can derive what are the conditions to satisfy the violation of the simplest example of the extension of the KCBS inequality in \cite{SZDM:pra16}. Our method can be systematically used in different experimental proposals as long as there is a theoretical model for the imprecision in the experiment. Our method could potentially be extended to the cases where there is no valid theoretical model and this problem could be addressed via tomography. This would be useful when the source of imperfection is hard to define correctly or to verify the theoretical model. Moreover, we also investigate how the decoherence can be taken into account together with the ontologically faithful non-contextuality tests.

While we had to make assumptions such as the ontologically faithful non-contextuality to address the challenges of the observation of contextuality we do not know how these assumptions affect the various advantages of contextuality in quantum information processing. Further investigation is needed in this direction.

\bigskip

\noindent {\bf Acknowledgements.}
AS has been supported by a KIAS individual grant (CG070301) at Korea Institute for Advanced Study. DM acknowledges funding from the ANR through
ANR-17-CE24-0035 VanQuTe.

\bibliography{biblio}

\onecolumngrid
\newpage
\appendix

\section{The Unitary Operations for the KCBS}\label{sec:appuniN5}

Based on the beam splitter matrices, we can compute the corresponding five unitary operators:

\begin{table}[ht]
\centering
\renewcommand{\arraystretch}{1.2}
\begin{tabular}{| c | c | c | c |}
\hline 
i & $\vert v_i \rangle$ & $U_i$ & Setup \\
\hline
1
& $\frac{1}{\sqrt{3}}
       \begin{pmatrix}
        1 \\
        -1\\
        1 \\
       \end{pmatrix}$
& $\frac{1}{\sqrt{6}}
       \begin{pmatrix}
        \sqrt{2} & -\sqrt{2} & \sqrt{2} \\
        2 & 1 & -1 \\
        0 & \sqrt{3} & \sqrt{3} \\
       \end{pmatrix}$
& \begin{minipage}{.25\textwidth}
  \centering
  \vspace{0.4cm}
  \includegraphics[width=20mm]{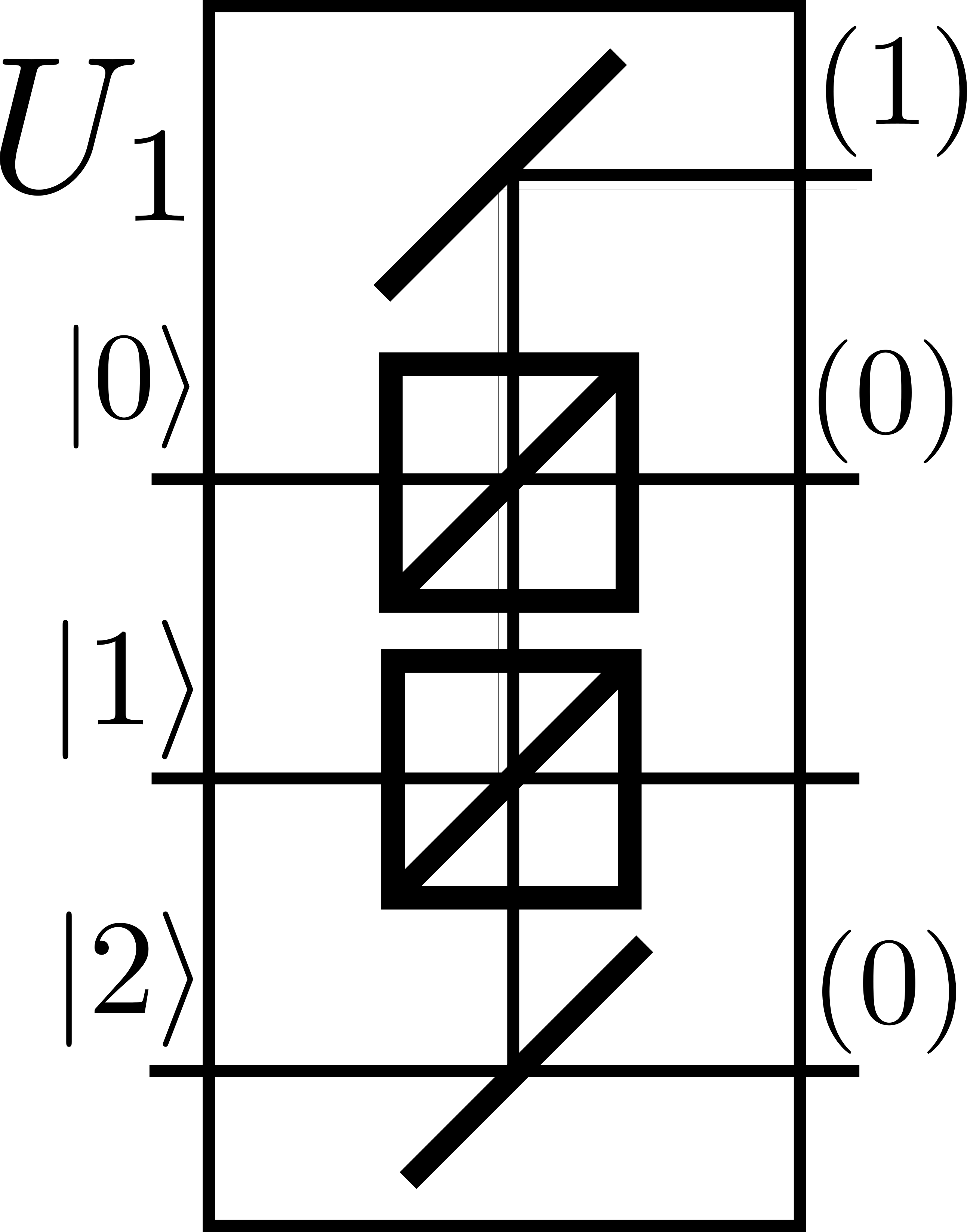}
  \vspace{0.4cm}
   \end{minipage}\\
\hline
2
& $\frac{1}{\sqrt{2}}
       \begin{pmatrix}
        1 \\
        1 \\
        0 \\

       \end{pmatrix}$
& $\frac{1}{\sqrt{2}}
          \begin{pmatrix}
          1 & 1 & 0 \\
          1 & -1 & 0 \\
          0 & 0 & \sqrt{2} \\
          \end{pmatrix}$
& \begin{minipage}{.25\textwidth}
  \centering
  \vspace{0.4cm}
  \includegraphics[width=20mm]{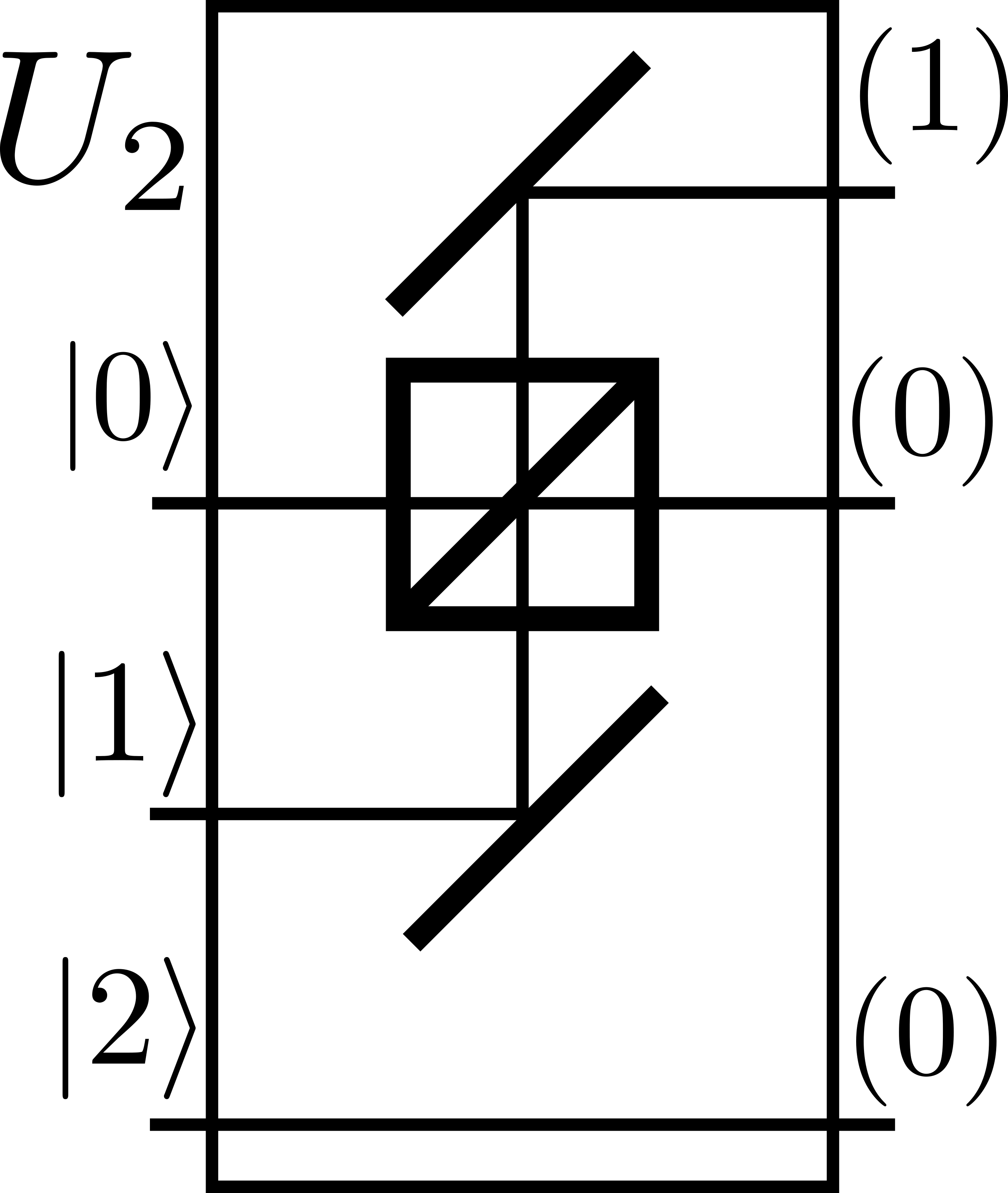}
  \vspace{0.4cm}
   \end{minipage}\\
\hline
3
& $
       \begin{pmatrix}
        0 \\
        0\\
        1 \\
       \end{pmatrix}$
& $ \begin{pmatrix}
         0 & 0 & 1 \\
         0 & 1 & 0 \\
         1 & 0 & 0 \\
         \end{pmatrix}$
& \begin{minipage}{.25\textwidth}
  \centering
  \vspace{0.4cm}
  \includegraphics[width=20mm]{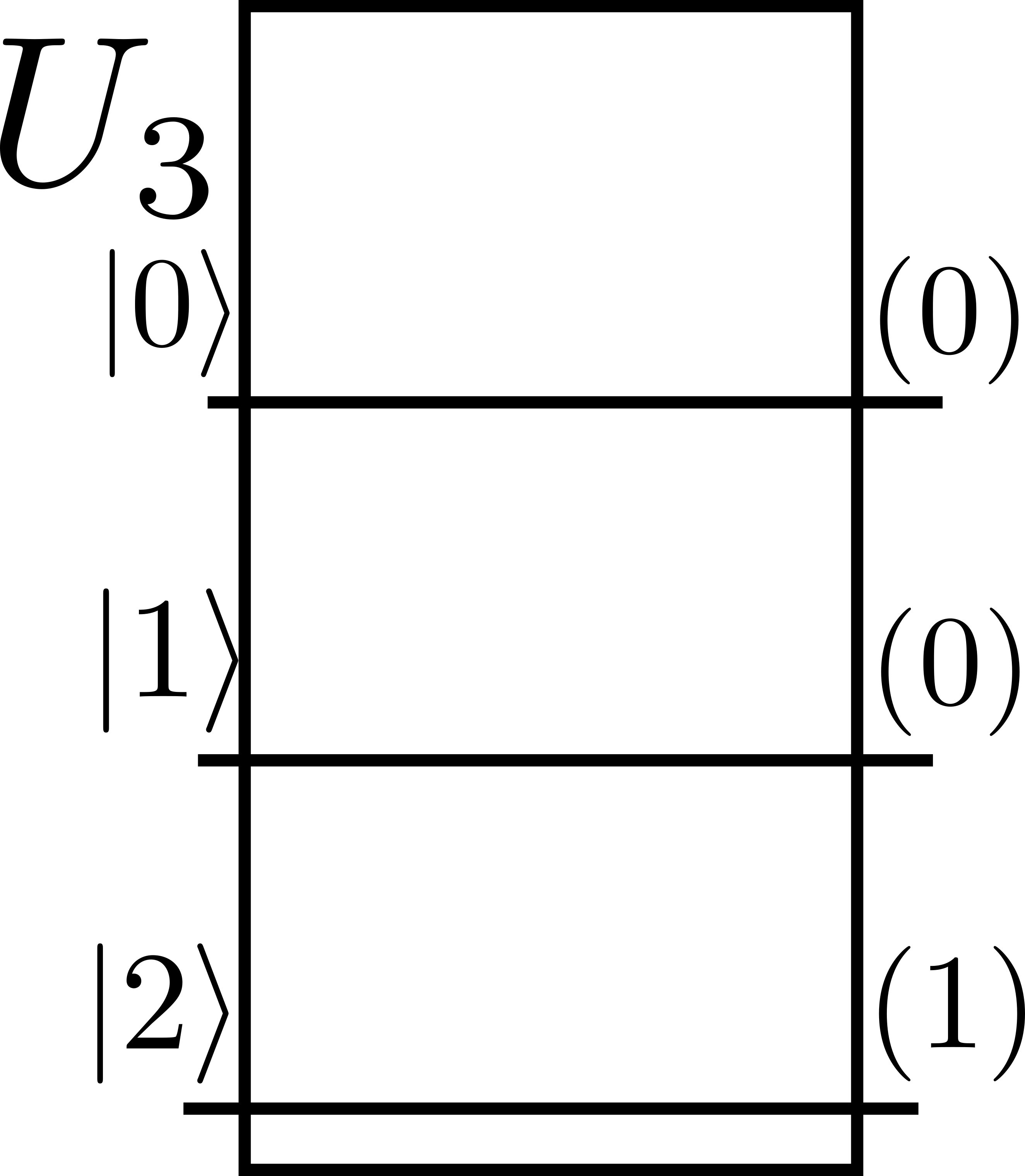}
  \vspace{0.4cm}
   \end{minipage}\\
\hline
4
& $
       \begin{pmatrix}
        1 \\
        0\\
        0 \\
       \end{pmatrix}$
& $ \begin{pmatrix}
         1 & 0 & 0 \\
         0 & 1 & 0 \\
         0 & 0 & 1 \\
         \end{pmatrix}$
&\begin{minipage}{.25\textwidth}
  \centering
  \vspace{0.4cm}
  \includegraphics[width=20mm]{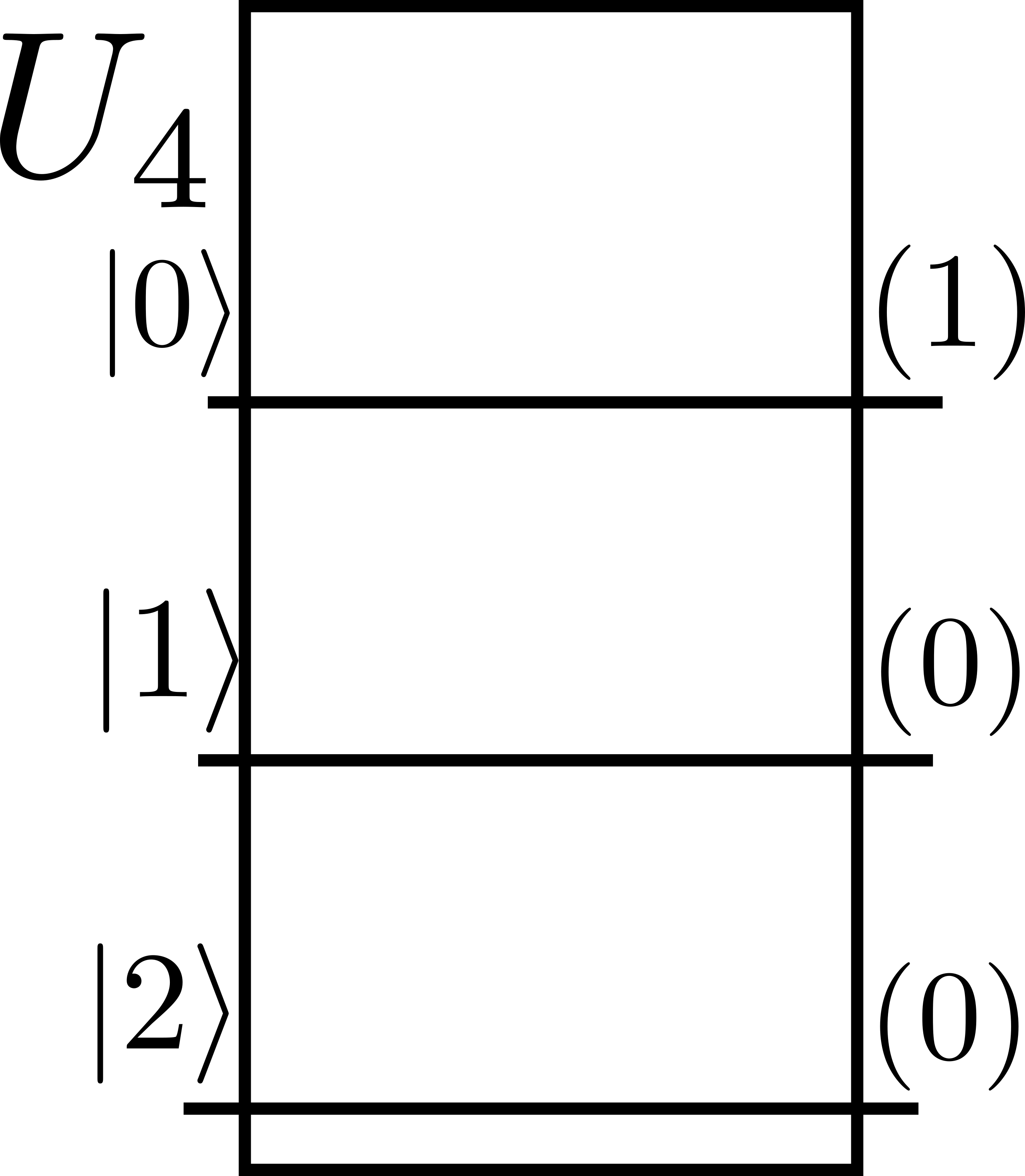}
  \vspace{0.4cm}
   \end{minipage}\\
\hline
5
& $\frac{1}{\sqrt{2}}
       \begin{pmatrix}
        0 \\
        1\\
        1 \\
       \end{pmatrix}$
& $\frac{1}{\sqrt{2}}
          \begin{pmatrix}
          0 & 1 & 1 \\
          0 & 1 & -1 \\
          \sqrt{2} & 0 & 0 \\
          \end{pmatrix}$
& \begin{minipage}{.25\textwidth}
  \centering
  \vspace{0.4cm}
  \includegraphics[width=20mm]{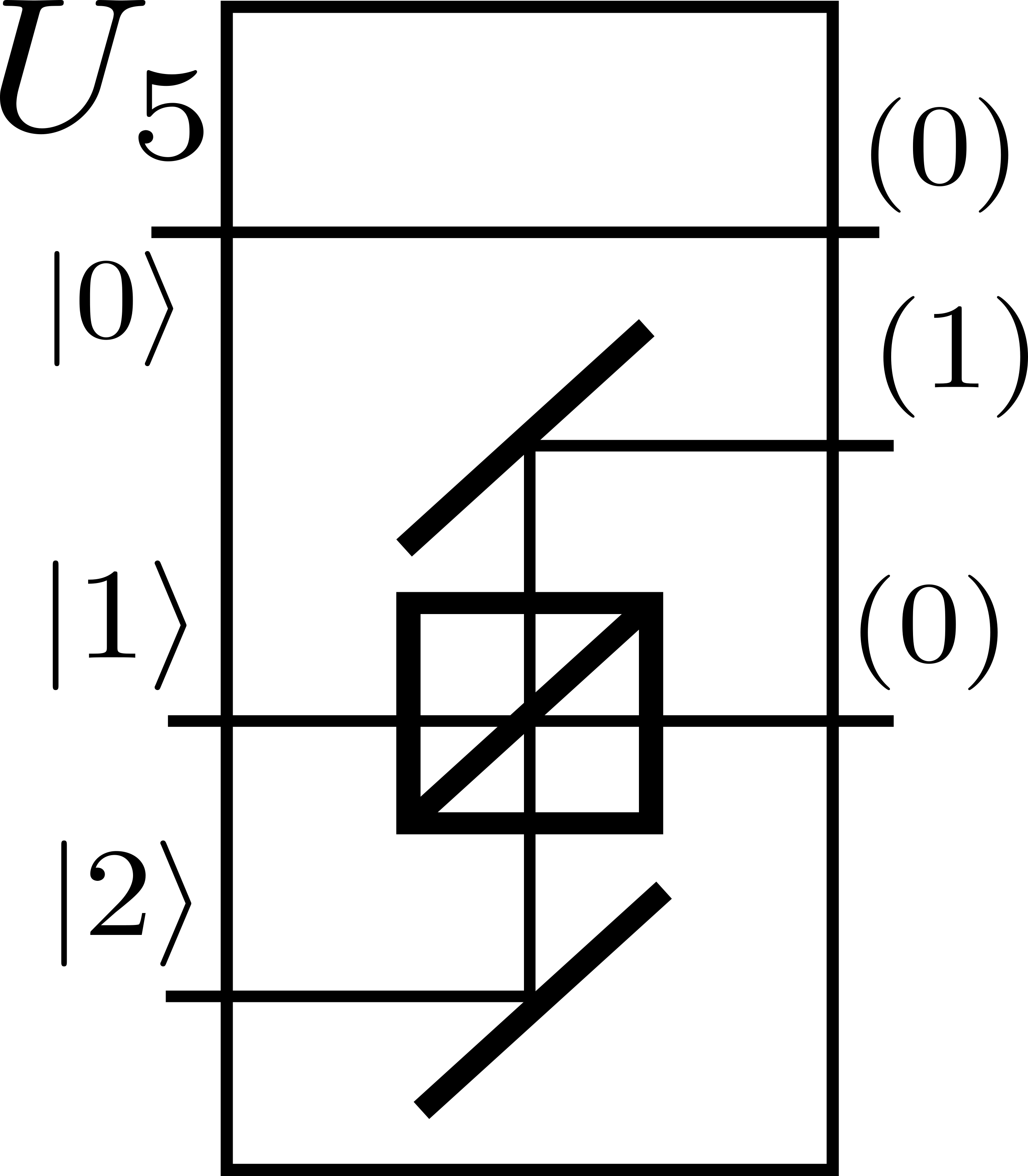}
  \vspace{0.4cm}
   \end{minipage}\\
\hline
\end{tabular}
\caption{Table showing for each vertex $i$, the value of $\vert v_i \rangle$, the matrix $U_i$ and the resulting optical network using beam splitters.}
\label{tab:expN5}
\end{table}

\subsection{Noisy Unitaries}\label{app:KCSBnoisyUi}

By adding the imperfection factor $\delta$ into a $50:50$ beam splitter as a dissymetry between the transmission and reflecion coefficients, we obtain :

\begin{equation}\label{eq:beamsplitterdelta}
 B_{\delta} = \frac{1}{\sqrt{2}}
 \begin{pmatrix}
 \sqrt{1+2\delta} & \sqrt{1-2\delta}\\
 \sqrt{1-2\delta} & - \sqrt{1+2\delta}\\
 \end{pmatrix},
\end{equation}
which gives for the reflection or transition the value $1/2\pm\delta$.
The five previous matrices become:

\begin{align}\label{eq:unitairepentadelta}
& U_1 = \frac{1}{\sqrt{6}}
    \begin{pmatrix}
     \sqrt{1+3(-1)^{\phi}\delta} & \sqrt{2-3(-1)^{\phi}\delta} & 0 \\
     \sqrt{2-3(-1)^{\phi}\delta} & \sqrt{1+3(-1)^{\phi}\delta} & 0 \\
     0 & 0 & \sqrt{3} \\
    \end{pmatrix}
    \text{ . }
    \begin{pmatrix}
    \sqrt{2} & 0 & 0 \\
    0 & -\sqrt{1+2\delta} & \sqrt{1-2\delta} \\
    0 & \sqrt{1-2\delta} & \sqrt{1+2\delta} \\
    \end{pmatrix},\notag\\
 & U_2 = \frac{1}{\sqrt{2}}
 \begin{pmatrix}
 \sqrt{1+2\delta} & \sqrt{1-2\delta} & 0 \\
 \sqrt{1-2\delta} & -\sqrt{1+2\delta} & 0 \\
 0 & 0 & \sqrt{2} \\
 \end{pmatrix},\notag\\
 & U_3 =
 \begin{pmatrix}
 0 & 0 & 1 \\
 0 & 1 & 0 \\
 1 & 0 & 0 \\
 \end{pmatrix},\notag\\
 & U_4 =
 \begin{pmatrix}
 1 & 0 & 0 \\
 0 & 1 & 0 \\
 0 & 0 & 1 \\
 \end{pmatrix},\notag\\
 & U_5 = \frac{1}{\sqrt{2}}
 \begin{pmatrix}
 0 & \sqrt{1+2\delta} & \sqrt{1-2\delta} \\
 0 & \sqrt{1-2\delta} & -\sqrt{1+2\delta} \\
 \sqrt{2} & 0 & 0 \\
 \end{pmatrix},
 \end{align}
where $\phi = \{0,1\}$, which permits when there are two beam splitters needed to consider the case where both imperfection factors change the reflection coefficient in the same way or in an opposite way.

\section{The Unitary Operations for $N=6$}
\subsection{Theoretical Unitaries}\label{app:N6Th}

\begin{align*}
  & U_1 =
 \frac{1}{\sqrt{3}}
 \begin{pmatrix}
     -1 & -\sqrt{2} & 0 & 0\\
     -\sqrt{2} & 1 & 0 & 0\\
     0 & 0 & \sqrt{3} & 0\\
     0 & 0 & 0 & \sqrt{3}\\
 \end{pmatrix}
 \text{ . } \frac{1}{2}
 \begin{pmatrix}
     2 & 0 & 0 & 0\\
     0 & -1 & \sqrt{3} & 0\\
     0 & \sqrt{3} & 1 & 0\\
     0 & 0 & 0 & 2\\
 \end{pmatrix}
  \text{ . } \frac{1}{\sqrt{3}}
  \begin{pmatrix}
     \sqrt{3} & 0 & 0 & 0\\
     0 & \sqrt{3} & 0 & 0\\
     0 & 0 & -1 & \sqrt{2}\\
     0 & 0 & \sqrt{2} & 1\\
  \end{pmatrix}\notag\\
 & U_ 1 = \frac{1}{6}
 \begin{pmatrix}
     -2\sqrt{3} & \sqrt{6} & \sqrt{6} & -2\sqrt{3}\\
     -2\sqrt{6} & -\sqrt{3} & -\sqrt{3} & \sqrt{6}\\
     0 & 3\sqrt{3} & -\sqrt{3} & \sqrt{6}\\
     0 & 0 & 2\sqrt{6} & 2\sqrt{3}\\
     \end{pmatrix},\notag\\
& U_2 =
     \begin{pmatrix}
      1 & 0 & 0 & 0\\
      0 & 1 & 0 & 0\\
      0 & 0 & 1 & 0\\
      0 & 0 & 0 & 1\\
     \end{pmatrix},\notag\\
& U_3 = \frac{1}{\sqrt{2}}
     \begin{pmatrix}
      1 & 1 & 0 & 0\\
      1 & -1 & 0 & 0\\
      0 & 0 & \sqrt{2} & 0\\
      0 & 0 & 0 & \sqrt{2}\\
      \end{pmatrix}
     \text{ . } \frac{1}{\sqrt{2}}
     \begin{pmatrix}
      0 & 0 & 0 & \sqrt{2}\\
      0 & 1 & 1 & 0\\
      0 & 1 & -1 & 0\\
      \sqrt{2} & 0 & 0 & 0\\
     \end{pmatrix}
     = \frac{1}{2}
     \begin{pmatrix}
      0 & 1 & 1 & \sqrt{2}\\
      0 & -1 & -1 & \sqrt{2}\\
      0 & \sqrt{2} & -\sqrt{2} & 0\\
      2 & 0 & 0 & 0\\
     \end{pmatrix},\notag \\
& U_4 = \frac{1}{\sqrt{2}}
    \begin{pmatrix}
     0 & 1 & 1 & 0\\
     0 & 1 & -1 & 0\\
     \sqrt{2} & 0 & 0 & 0\\
     0 & 0 & 0 & \sqrt{2}\\
    \end{pmatrix},\notag\\
 \end{align*}

 \begin{align}
& U_5 = \frac{1}{\sqrt{2}}
     \begin{pmatrix}
      1 & 1 & 0 & 0\\
      -1 & 1 & 0 & 0\\
      0 & 0 & \sqrt{2} & 0\\
      0 & 0 & 0 & \sqrt{2}\\
     \end{pmatrix}
     \text{ . } \frac{1}{\sqrt{2}}
     \begin{pmatrix}
      \sqrt{2} & 0 & 0 & 0\\
      0 & 1 & 1 & 0\\
      0 & 1 & -1 & 0\\
      0 & 0 & 0 & \sqrt{2}\\
     \end{pmatrix}
     = \frac{1}{2}
     \begin{pmatrix}
      \sqrt{2} & 1 & 1 & 0\\
      -\sqrt{2} & 1 & 1 & 0\\
      0 & \sqrt{2} & -\sqrt{2} & 0\\
      0 & 0 & 0 & 2\\
     \end{pmatrix},\notag\\
& U_6 =
     \begin{pmatrix}
      0 & 0 & 0 & 1\\
      0 & 1 & 0 & 0\\
      0 & 0 & 1 & 0\\
      1 & 0 & 0 & 0\\
     \end{pmatrix}.
\end{align}

We represent the setup composed by beam splitters in the following Tab.~\cref{tab:expN6}.

\begin{table}[ht]
\centering
\renewcommand{\arraystretch}{1.2}
\begin{tabular}{| c | c | c | c |}
\hline 
i & $\vert v_i \rangle$ & $U_i$ & Setup \\
\hline
1
& $\frac{1}{\sqrt{6}}
       \begin{pmatrix}
        -\sqrt{2} \\
        1 \\
        1 \\
        -\sqrt{2}
       \end{pmatrix}$
& $\frac{1}{6}
 \begin{pmatrix}
     -2\sqrt{3} & \sqrt{6} & \sqrt{6} & -2\sqrt{3}\\
     -2\sqrt{6} & -\sqrt{3} & -\sqrt{3} & \sqrt{6}\\
     0 & 3\sqrt{3} & -\sqrt{3} & \sqrt{6}\\
     0 & 0 & 2\sqrt{6} & 2\sqrt{3}\\
     \end{pmatrix}$
& \begin{minipage}{.25\textwidth}
  \centering
  \vspace{0.4cm}
  \includegraphics[width=20mm]{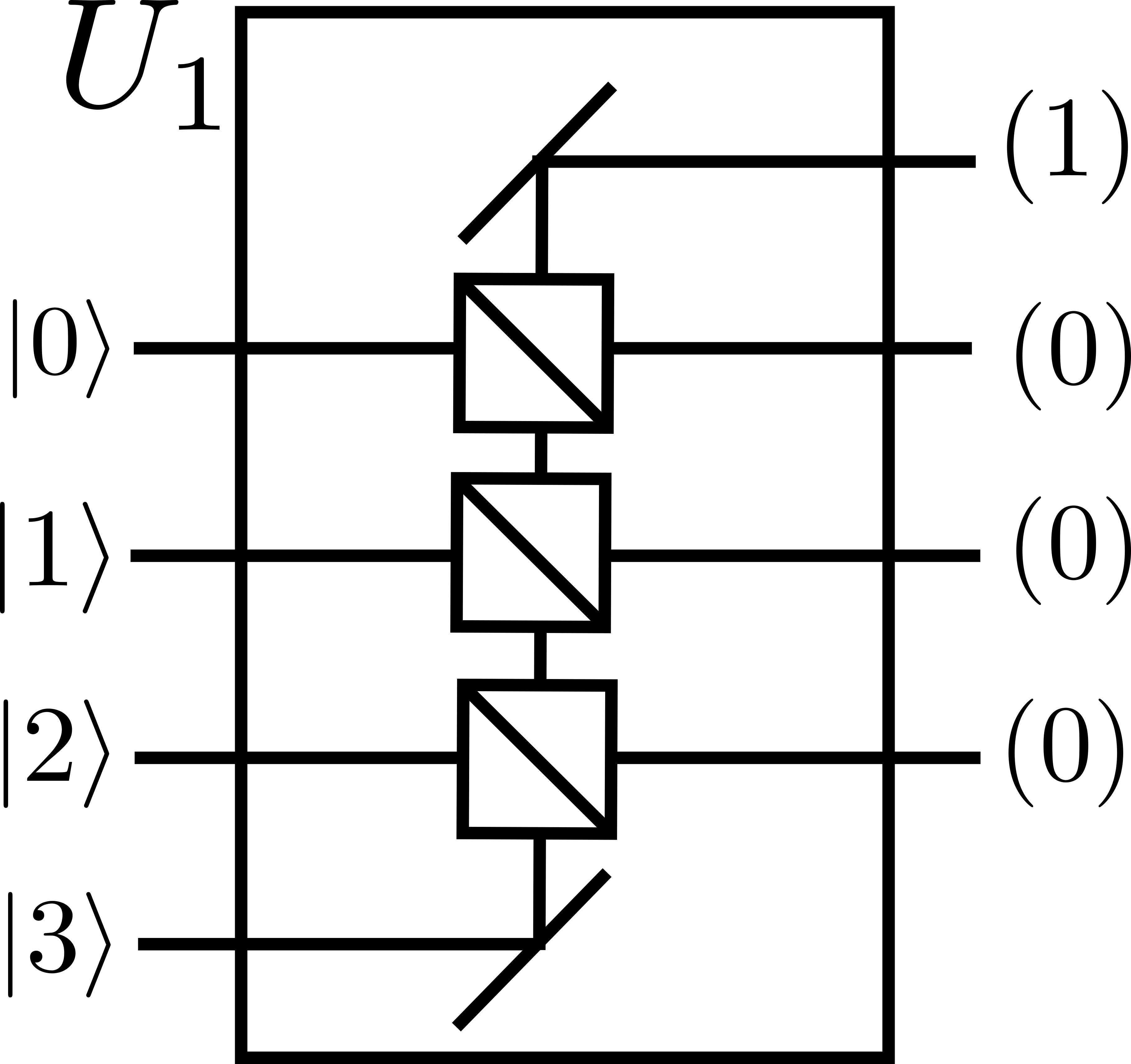}
  \vspace{0.4cm}
   \end{minipage}\\
\hline
2
& $
       \begin{pmatrix}
        1 \\
        0\\
        0 \\
        0\\
       \end{pmatrix}$
& $\begin{pmatrix}
      1 & 0 & 0 & 0\\
      0 & 1 & 0 & 0\\
      0 & 0 & 1 & 0\\
      0 & 0 & 0 & 1\\
     \end{pmatrix}$
& \begin{minipage}{.25\textwidth}
  \centering
  \vspace{0.4cm}
  \includegraphics[width=20mm]{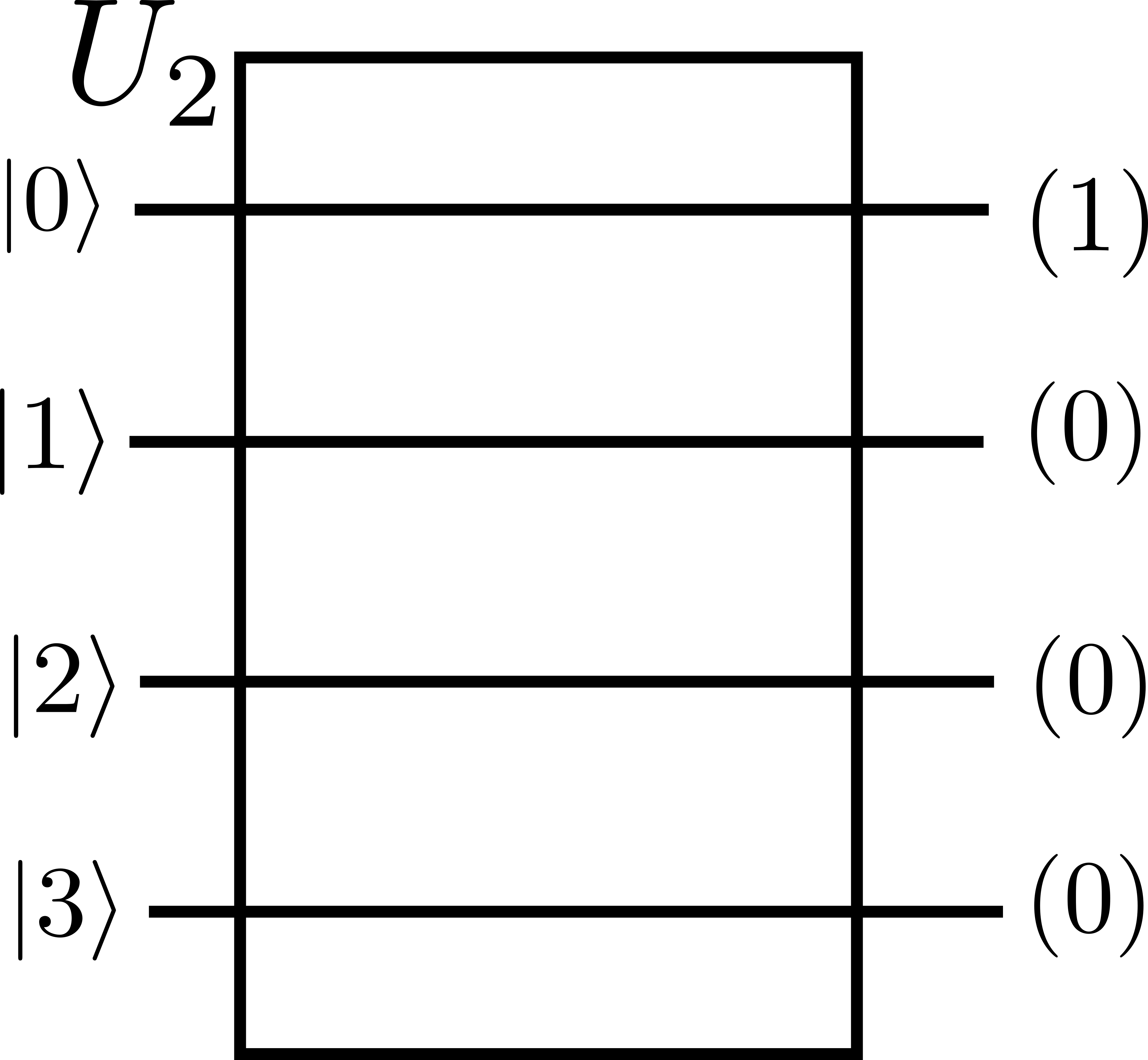}
  \vspace{0.4cm}
   \end{minipage}\\
\hline
3
& $ \frac{1}{2}
       \begin{pmatrix}
        0 \\
        1\\
        1 \\
        \sqrt{2}\\
       \end{pmatrix}$
& $ \frac{1}{2}
     \begin{pmatrix}
      0 & 1 & 1 & \sqrt{2}\\
      0 & -1 & -1 & \sqrt{2}\\
      0 & \sqrt{2} & -\sqrt{2} & 0\\
      2 & 0 & 0 & 0\\
     \end{pmatrix}$
& \begin{minipage}{.25\textwidth}
  \centering
  \vspace{0.4cm}
  \includegraphics[width=20mm]{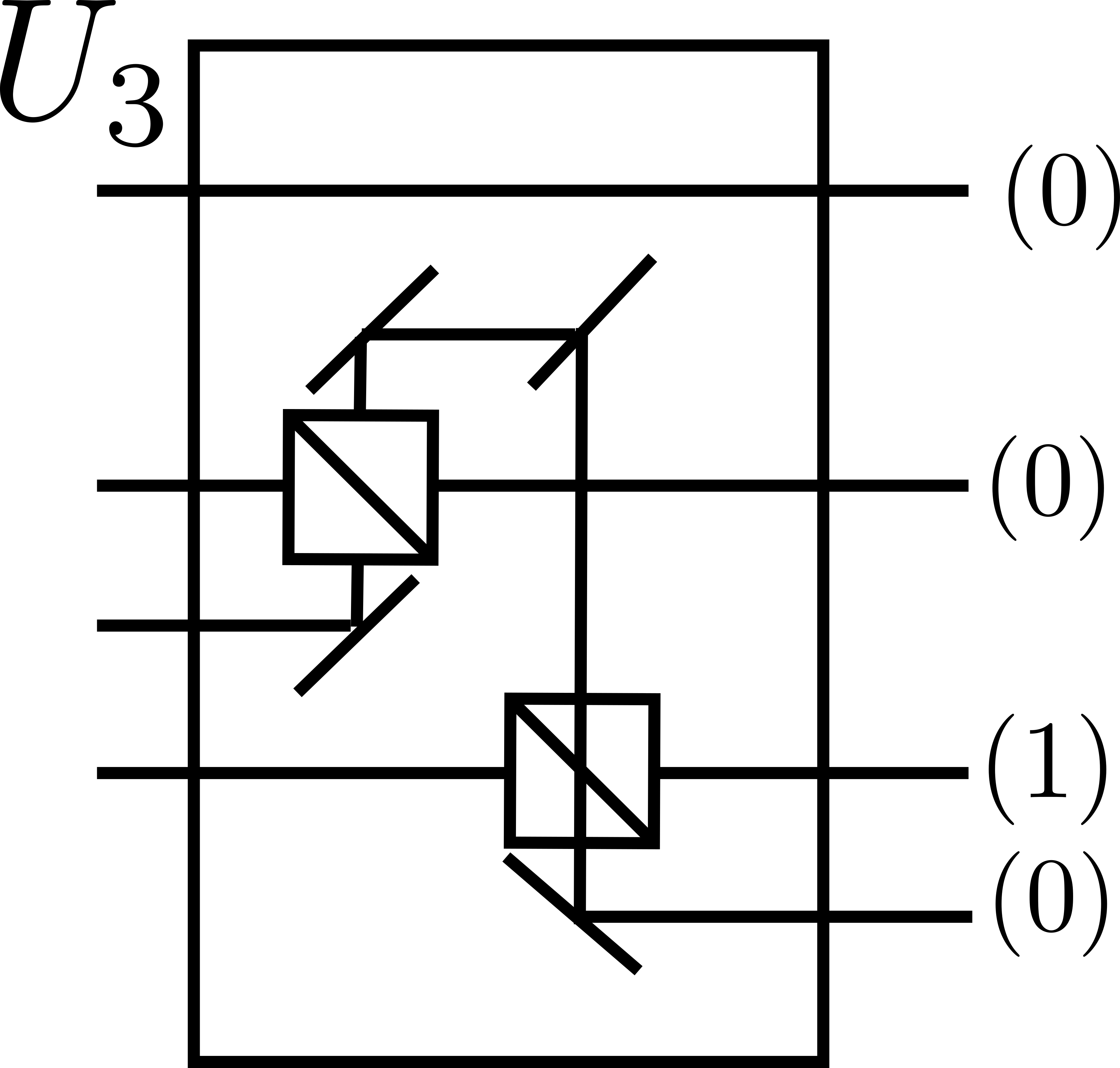}
  \vspace{0.4cm}
   \end{minipage}\\
\hline
4
& $ \frac{1}{\sqrt{2}}
       \begin{pmatrix}
        0 \\
        1\\
        1 \\
        0\\
       \end{pmatrix}$
& $ \frac{1}{\sqrt{2}}
          \begin{pmatrix}
           0 & 1 & 1 & 0\\
           0 & 1 & -1 & 0\\
           \sqrt{2} & 0 & 0 & 0\\
           0 & 0 & 0 & \sqrt{2}\\
          \end{pmatrix}$
&\begin{minipage}{.25\textwidth}
  \centering
  \vspace{0.4cm}
  \includegraphics[width=20mm]{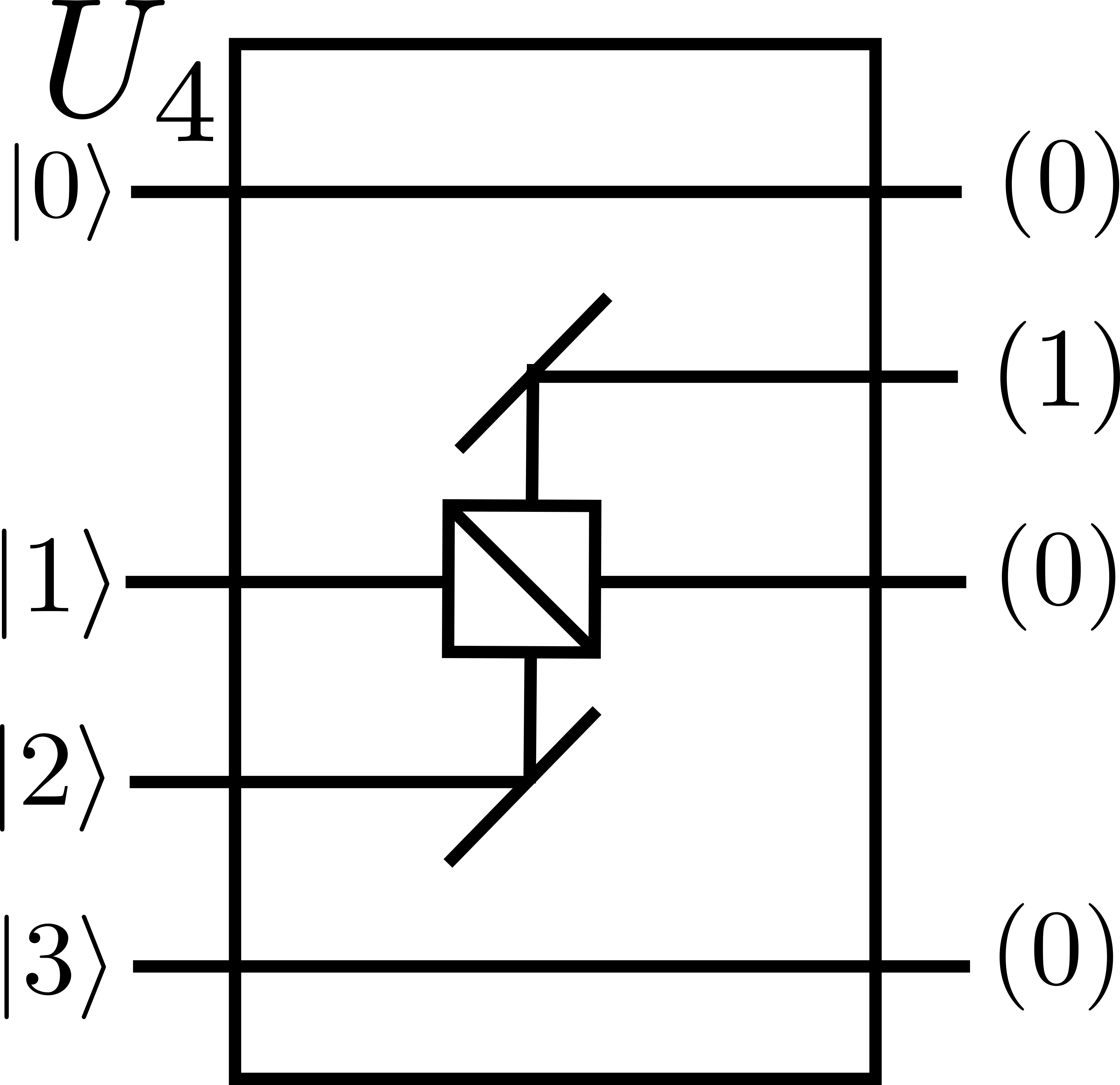}
  \vspace{0.4cm}
   \end{minipage}\\
\hline
5
& $\frac{1}{2}
       \begin{pmatrix}
        \sqrt{2} \\
        1\\
        1 \\
        0\\
       \end{pmatrix}$
& $\frac{1}{2}
     \begin{pmatrix}
      \sqrt{2} & 1 & 1 & 0\\
      -\sqrt{2} & 1 & 1 & 0\\
      0 & \sqrt{2} & -\sqrt{2} & 0\\
      0 & 0 & 0 & 2\\
     \end{pmatrix}$
& \begin{minipage}{.25\textwidth}
  \centering
  \vspace{0.4cm}
  \includegraphics[width=20mm]{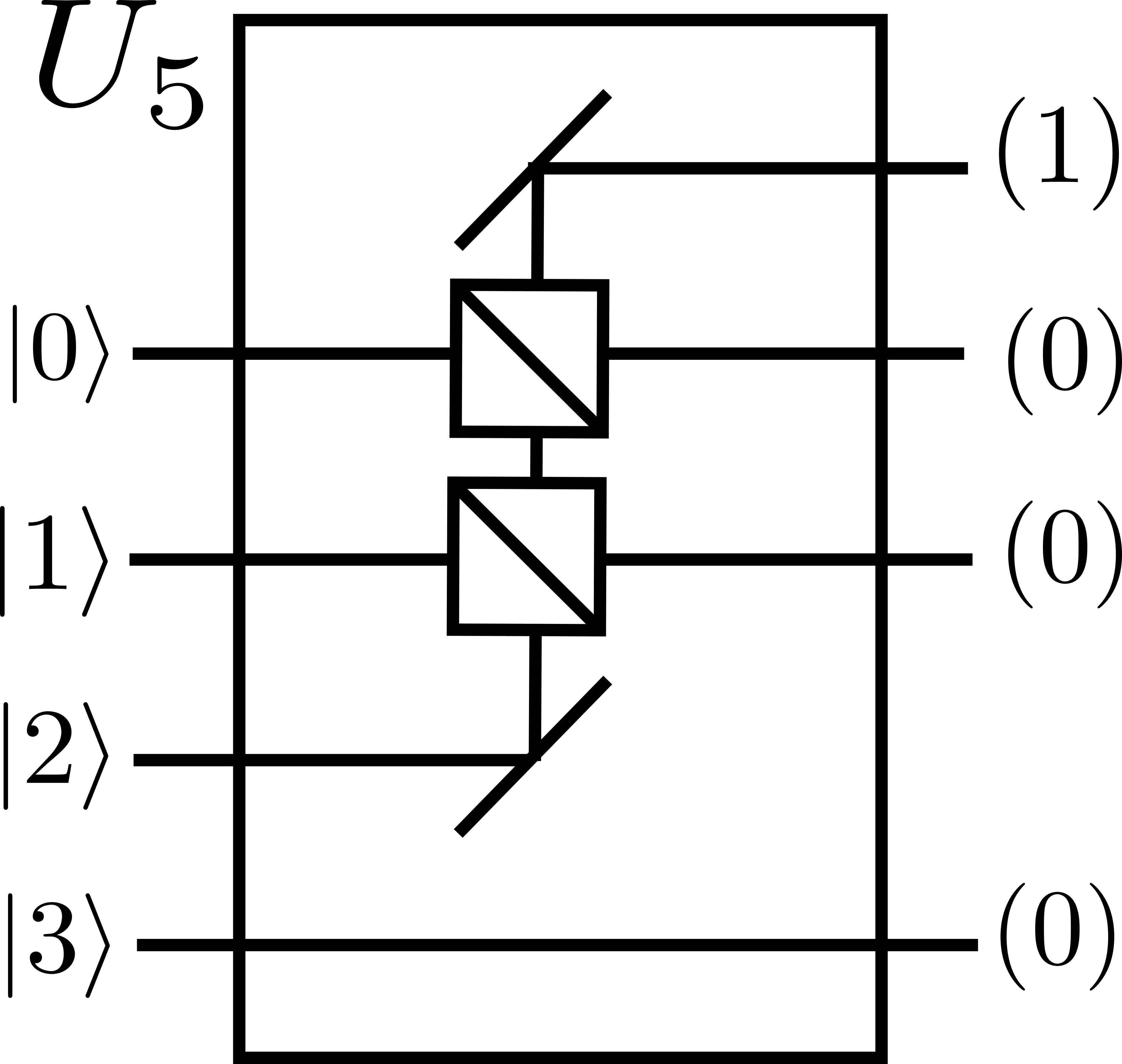}
  \vspace{0.4cm}
   \end{minipage}\\
\hline
6
& $
       \begin{pmatrix}
        0 \\
        0\\
        0\\
        1 \\
       \end{pmatrix}$
& $\begin{pmatrix}
      0 & 0 & 0 & 1\\
      0 & 1 & 0 & 0\\
      0 & 0 & 1 & 0\\
      1 & 0 & 0 & 0\\
     \end{pmatrix}$
& \begin{minipage}{.25\textwidth}
  \centering
  \vspace{0.4cm}
  \includegraphics[width=20mm]{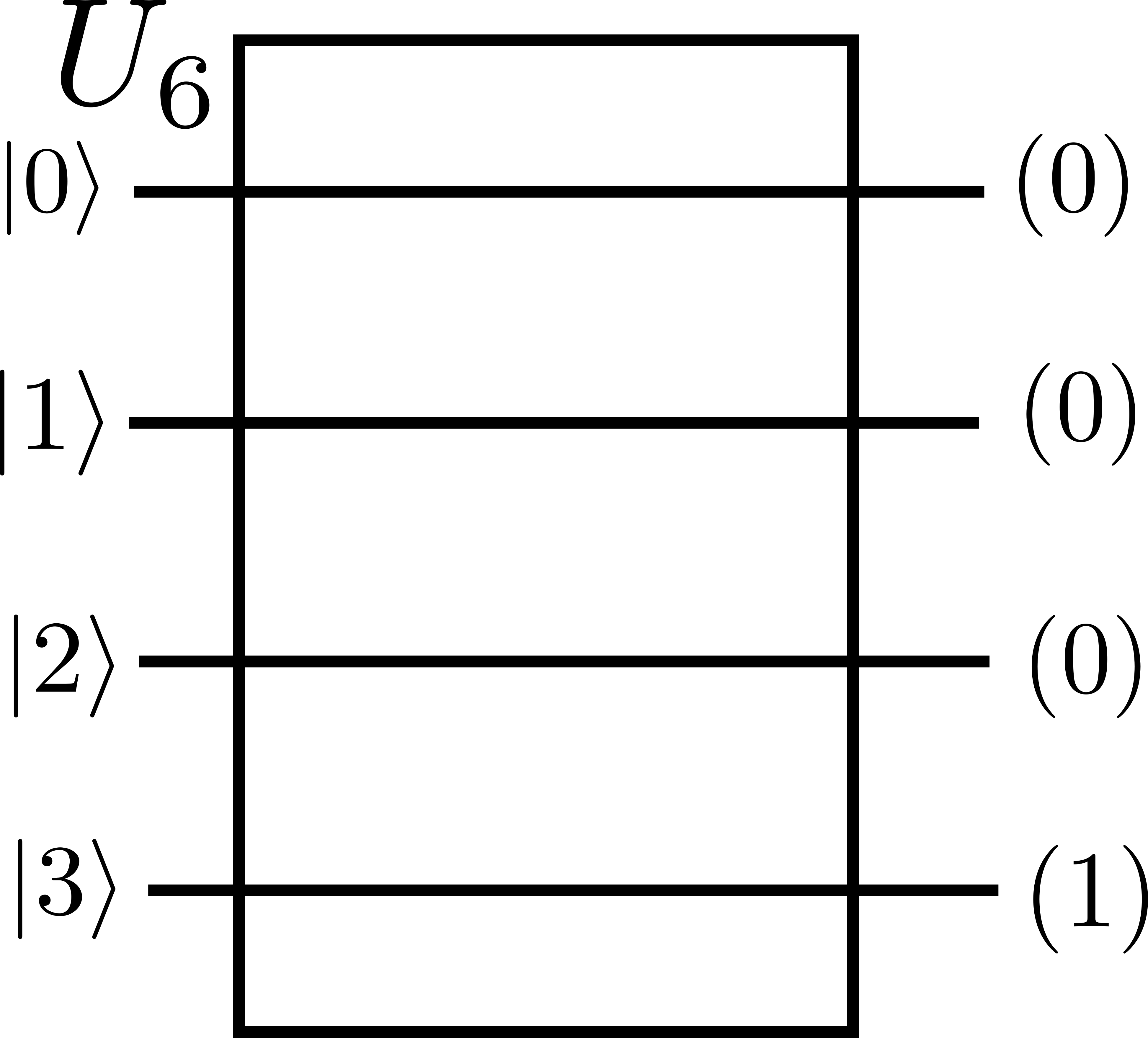}
  \vspace{0.4cm}
   \end{minipage}\\
\hline
\end{tabular}
\caption{Table showing for each vertex $i$, the value of $\vert v_i \rangle$, the matrix $U_i$ and the resulting optical network using beam splitters.}
\label{tab:expN6}
\end{table}

\subsection{Noisy Unitary for $N=6$}

\begin{align*}
& U_1 =
\frac{1}{\sqrt{3}}
\begin{pmatrix}
    -\sqrt{1+3(-1)^{\phi_2}\delta} & -\sqrt{2-3(-1)^{\phi_2}\delta} & 0 & 0\\
    -\sqrt{2-3(-1)^{\phi_2}\delta} & \sqrt{1+3(-1)^{\phi_2}\delta} & 0 & 0\\
    0 & 0 & \sqrt{3} & 0\\
    0 & 0 & 0 & \sqrt{3}\\
\end{pmatrix}\notag\\
& \text{ . } \frac{1}{2}
\begin{pmatrix}
    1 & 0 & 0 & 0\\
    0 & -\sqrt{1+4(-1)^{\phi_1}\delta} & \sqrt{3-4(-1)^{\phi_1}\delta} & 0\\
    0 & \sqrt{3-4(-1)^{\phi_1}\delta} & \sqrt{1+4(-1)^{\phi_1}\delta} & 0\\
    0 & 0 & 0 & 1\\
\end{pmatrix}
 \text{ . } \frac{1}{\sqrt{3}}
 \begin{pmatrix}
    \sqrt{3} & 0 & 0 & 0\\
    0 & \sqrt{3} & 0 & 0\\
    0 & 0 & -\sqrt{1+3\delta} & \sqrt{2+3\delta}\\
    0 & 0 & \sqrt{2+3\delta} & \sqrt{1+3\delta}\\
 \end{pmatrix},\notag\\
 & U_2 =
\begin{pmatrix}
    1 & 0 & 0 & 0\\
    0 & 1 & 0 & 0\\
    0 & 0 & 1 & 0\\
    0 & 0 & 0 & 1\\
\end{pmatrix},\notag\\
& U_3 = \frac{1}{\sqrt{2}}
\begin{pmatrix}
    \sqrt{1+2\delta} & \sqrt{1-2\delta} & 0 & 0\\
    \sqrt{1-2\delta} & -\sqrt{1+2\delta} & 0 & 0\\
    0 & 0 & \sqrt{2} & 0\\
    0 & 0 & 0 & \sqrt{2}\\
\end{pmatrix}\notag\\
& \text{ . } \frac{1}{\sqrt{2}}
\begin{pmatrix}
    0 & 0 & 0 & \sqrt{2}\\
    0 & \sqrt{1+2(-1)^{\phi}\delta} & \sqrt{1-2(-1)^{\phi}\delta} & 0\\
    0 & \sqrt{1-2(-1)^{\phi}\delta} & -\sqrt{1+2(-1)^{\phi}\delta} & 0\\
    \sqrt{2} & 0 & 0 & 0\\
\end{pmatrix},\notag\\
& U_4 = \frac{1}{\sqrt{2}}
\begin{pmatrix}
    0 & -\sqrt{1+2\delta} & \sqrt{1-2\delta} & 0\\
    0 & \sqrt{1-2\delta} & \sqrt{1+2\delta} & 0\\
    \sqrt{2} & 0 & 0 & 0\\
    0 & 0 & 0 & \sqrt{2}\\
\end{pmatrix},\notag\\
\end{align*}
\begin{align}
& U_5 = \frac{1}{\sqrt{2}}
\begin{pmatrix}
    \sqrt{1+2(-1)^{\phi}\delta} & \sqrt{1-2(-1)^{\phi}\delta} & 0 & 0\\
    -\sqrt{1-2(-1)^{\phi}\delta} & \sqrt{1+2(-1)^{\phi}\delta} & 0 & 0\\
    0 & 0 & \sqrt{2} & 0\\
    0 & 0 & 0 & \sqrt{2}\\
\end{pmatrix} \text{ . } \frac{1}{\sqrt{2}}
\begin{pmatrix}
    \sqrt{2} & 0 & 0 & 0\\
    0 & \sqrt{1+2\delta} & \sqrt{1-2\delta} & 0\\
    0 & \sqrt{1-2\delta} & -\sqrt{1+2\delta} & 0\\
    0 & 0 & 0 & \sqrt{2}\\
\end{pmatrix},\notag\\
& U_6 =
\begin{pmatrix}
    0 & 0 & 0 & 1\\
    0 & 1 & 0 & 0\\
    0 & 0 & 1 & 0\\
    1 & 0 & 0 & 0\\
\end{pmatrix}.
\end{align}
where $\phi$ , $\phi_1$ and $\phi_2 \in \{0,1\}$.

\end{document}